
 \input harvmac

\font\cmss=cmss10 \font\cmsss=cmss10 at 7pt
\def\IZ{\relax\ifmmode\mathchoice
{\hbox{\cmss Z\kern-.4em Z}}{\hbox{\cmss Z\kern-.4em Z}}
{\lower.9pt\hbox{\cmsss Z\kern-.4em Z}}
{\lower1.2pt\hbox{\cmsss Z\kern-.4em Z}}\else{\cmss Z\kern-.4em Z}\fi}

\vsize=8.75truein
\hsize=5.5truein
\voffset=0.25truein

\nopagenumbers

\Title{\vbox{\baselineskip12pt\hbox{FTUAM-26/93}\hbox{}}}
{\vbox{\centerline{Towards a Theory of Soft Terms}
   \vskip2pt\centerline{for the Supersymmetric Standard Model }}}

\centerline{ \bf A. Brignole, L.E. Ib\'a\~nez and C. Mu\~noz }
\centerline{ Departamento de  F\'{\i}sica Te\'orica C-XI}
\centerline{ Universidad Aut\'onoma de Madrid}
\centerline{ Cantoblanco,
28049, Madrid, Spain}

 \vskip .2in
 \noindent
\centerline{\bf ABSTRACT}
\bigskip
\bigskip
\bigskip

\vbox{\baselineskip12pt We perform a systematic analysis of the  soft
supersymmetry-breaking terms arising
in some large classes of four-dimensional strings.
The analysis does not assume any specific supersymmetry-breaking
mechanism but provides a means of parametrizing our ignorance in
a way consistent with some known properties of these four-dimensional
strings. We introduce a     {\it goldstino angle}
parameter $\theta $ which says where the source
 of supersymmetry-breaking resides,
either  predominantly in the
dilaton sector ($sin \theta =1$ limit) or in the rest of the chiral fieds,
notably the moduli ($sin\theta =0$ limit). All formulae for soft
parameters take particularly simple forms when written in terms of this angle.
The $sin\theta =1$ limit is (up to small corrections) universal.
As $sin\theta $ decreases,
the model dependence increases and the resulting soft terms may
or may not be universal, depending on the model.
General expressions for the soft terms as functions of $\theta$
 for generic four-dimensional strings are provided.
For each {\it given} string model, one trades the four soft
parameters ($M,m,A,B$) of the minimal supersymmetric standard model
by the two parameters $m_{3/2}$ (gravitino mass) and $sin\theta $.
 The role of complex
phases and the associated constraints
from limits on  the electric dipole moment of the neutron are  discussed.
 It is also emphasized the importance of treating the problem
of the cosmological constant in a self-consistent manner.
Three prototype string scenarios are discussed and
                             their low-energy implications are studied by
imposing appropriate radiative $SU(2)_L\times U(1)$ breaking. The
supersymmetric
particle spectra present definite patterns which may be experimentally tested
at future colliders.
}

\Date{August 1993}
 \noblackbox

\lref\IN{L.E. Ib\'{a}\~{n}ez and H.P. Nilles, Phys. Lett. 169B
(1986) 354}
\lref\DFKZ{J.P. Derendinger, S. Ferrara, C. Kounnas and F. Zwirner,
Nucl. Phys. B372 (1992) 145, Phys. Lett. B271 (1991) 307}
\lref\W{E. Witten, Phys. Lett. B155 (1985) 151}
\lref\CFGP{E. Cremmer, S. Ferrara, L. Girardello and A. Van
Proeyen, Nucl. Phys. B212 (1983) 413}
\lref\CFILQ{M. Cveti\u{c}, A. Font, L.E. Ib\'{a}\~{n}ez, D. L\"{u}st and
F. Quevedo, Nucl. Phys. B361 (1991) 194}
\lref\IL{L.E. Ib\'{a}\~{n}ez and D. L\"{u}st, Nucl. Phys. B382
(1992) 305}
\lref\CCM{B. de Carlos, J.A. Casas and C. Mu\~{n}oz, Phys. Lett.
B299 (1993) 234}
\lref\KL{V. Kaplunovsky and J. Louis, Phys. Lett. B306 (1993) 269}
\lref\SW{S.K. Soni and H.A. Weldon, Phys. Lett. 126B (1983) 215}
\lref\CD{L. Dixon, unpublished}
\lref\C{S. Ferrara, C. Kounnas, D. L\"ust and F. Zwirner,
Nucl. Phys. B365 (1991) 431;
M. Cveti\u{c}, talk given at the International Conference on
High Energy Physics, Dallas (1992)}
\lref\NSW{H.P. Nilles, M. Srednicki and D. Wyler, Phys. Lett.
B120 (1983) 346; J.M. Frere, D.R.T. Jones and S. Raby, Nucl. Phys.
B222 (1983) 11; J.P. Derendinger and C. Savoy, Nucl. Phys. B237
(1984) 307; L. E. Ib\'a\~nez and J. Mas, Nucl. Phys.B286 (1987) 107;
 J. Ellis, J.F. Gunion, H.E. Haber, L. Roszkowski and
F. Zwirner, Phys. Rev. D39 (1989) 844; M. Drees, Int. J. Mod. Phys.
A4 (1989) 3635}
\lref\KN{J.E. Kim and H.P. Nilles, Phys. Lett. B138 (1984) 150;
J.E. Kim
and H.P. Nilles, Phys. Lett. B263 (1991) 79; E.J. Chun, J.E. Kim and
H.P. Nilles, Nucl. Phys. B370 (1992) 105}
\lref\MUCM{J.A. Casas and C. Mu\~{n}oz,
Phys. Lett. B306 (1993) 288}
\lref\GM{G.F. Giudice and A. Masiero, Phys. Lett. B206 (1988) 480}
\lref\HLW{L. Hall, J. Lykken and S. Weinberg, Phys. Rev. D27 (1983) 2359}
\lref\CHSW{P. Candelas, G. Horowitz, A. Strominger and E. Witten,
Nucl. Phys. B258 (1985) 46}
\lref\DHVW{L. Dixon, J. Harvey, C. Vafa and E. Witten, Nucl.
Phys. B261 (1985) 651, B274 (1986) 285}
\lref\FKP{S. Ferrara,
C. Kounnas and M. Porrati, Phys. Lett. B181 (1986) 263; M. Cveti\u{c},
J. Louis and B. Ovrut, Phys. Lett. B206 (1988) 227}
\lref\DKL{L. Dixon, V. Kaplunovsky and J. Louis, Nucl. Phys.
B329 (1990) 27}
\lref\FLT{S. Ferrara, D. L\"{u}st A. Shapere and S. Theisen, Phys. Lett.
B225 (1989) 363; S. Ferrara, D. L\"{u}st and S. Theisen, Phys. Lett.
B233 (1989) 147}
\lref\HV{S. Hamidi and C. Vafa, Nucl. Phys. B279 (1987) 465;
L. Dixon, D. Friedan, E. Martinec and S. Shenker, Nucl. Phys. B282
(1987) 13}
\lref\CM{J.A. Casas and C. Mu\~{n}oz, Nucl. Phys. B332 (1990) 189;
J.A. Casas, F. G\'{o}mez and C. Mu\~{n}oz, Phys. Lett. B292 (1992) 42}
\lref\K{V. Kaplunovsky, Nucl. Phys. B307 (1988) 145 and erratum (1992)}
\lref\DKLD{L. Dixon,
V. Kaplunovsky and J. Louis, Nucl. Phys. B355 (1991) 649}
\lref\L{J. Louis, talk given at the 2nd International
Symposium on Particles, Strings and Cosmology, Boston (1991)}
\lref\LO{G. Lopes Cardoso and B. Ovrut, Nucl. Phys. B369 (1992) 351}
\lref\DGH{M. Dugan, B. Grinstein and L. Hall, Nucl. Phys. B255 (1985) 413}
\lref\ILCP{L.E. Ib\'{a}\~{n}ez and D. L\"{u}st, Phys. Lett. B267
(1991) 51}
\lref\FILQS{A. Font, L.E. Ib\'{a}\~{n}ez and D. L\"{u}st and
F. Quevedo, Phys. Lett. B249 (1990) 35}
\lref\FILQ{A. Font, L.E. Ib\'{a}\~{n}ez, D. L\"{u}st and F. Quevedo,
Phys. Lett. B245 (1990) 401}
\lref\FMTV{S. Ferrara, N. Magnoli, T.R. Taylor and G. Veneziano, Phys.
Lett. B245 (1990) 409}
\lref\A{N.V. Krasnikov, Phys. Lett. B193 (1987) 37; L. Dixon,
talk presented at the A.P.S. D.P.F. Meeting
at Houston (1990); V. Kaplunovsky, talk presented at the "Strings 90"
workshop at College Station (1990); J.A. Casas, Z. Lalak, C. Mu\~{n}oz
and G.G. Ross, Nucl. Phys. B347 (1990) 243; H.P. Nilles and M. Olechowsky,
Phys. Lett. B248 (1990) 268; P. Bin\'{e}truy and M.K. Gaillard,
Phys. Lett. B253 (1991) 119; D. L\"{u}st and T.R. Taylor,
Phys. Lett. B253 (1991) 335; B. de Carlos, J.A. Casas and C. Mu\~{n}oz,
Phys. Lett. B263 (1991) 248; J. Louis, talk presented at the Particles and
Fields'91 Symposium, Vancouver (1991); D. L\"{u}st and C. Mu\~{n}oz, Phys.
Lett. B279 (1992) 272}
\lref\ILR{L.E. Ib\'{a}\~{n}ez, D. L\"{u}st and G.G. Ross,
Phys. Lett. B272 (1991) 251}
\lref\AEKN{I. Antoniadis, J. Ellis, S. Kelley and D.V. Nanopoulos,
Phys. Lett. B271 (1991) 31}

\lref\ILM{L.E. Ib\'{a}\~{n}ez, C. L\'{o}pez and C. Mu\~{n}oz,
Nucl. Phys. B256 (1985) 218}
\lref\ILO{L.E. Ib\'{a}\~{n}ez and C. L\'{o}pez,
Nucl. Phys. B233 (1984) 511}
\lref\R{L.E. Ib\'{a}\~{n}ez and G.G. Ross, CERN--TH.6412/92 (1992),
to appear in Perspectives in Higgs Physics, ed. G. Kane}
\lref\CCMD{B. de Carlos, J.A. Casas and C. Mu\~{n}oz,
 Nucl. Phys. B399 (1993) 623}
\lref\BLM{R. Barbieri, J. Louis and M. Moretti, CERN--TH.6856/93
(1993) revised version}
\lref\gcon{J.P. Derendinger, L.E. Ib\'a\~nez and H.P. Nilles,
Phys. Lett. B155 (1985) 65; M. Dine, R. Rohm, N. Seiberg and E. Witten,
Phys. Lett. B156 (1985) 55}
\lref\DGH{M. Dugan, B. Grinstein and L. Hall, Nucl. Phys. B255 (1985) 413}
\lref\HAG{J. Hagelin, S. Kelley and T. Tanaka, MIU-THP-92/60, (1992)}
\lref\BC{B. Campbell, Phys.Rev. D28 (1983) 209 and references
therein}
\lref\DIM{J. L\'opez, D. Nanopoulos and A. Zichichi, CERN-TH-6926-93
(1993) revised version}
\lref\PX{P. Candelas, X. De la Ossa, P. Green et al.,
Nucl. Phys.B359 (1991) 21}
\lref\AMQ{I. Antoniadis, C. Mu\~noz and M. Quir\'os,
Nucl. Phys. B397 (1993) 515}
\lref\MS{ P. Mayr and S. Stieberger, Munich preprint MPI-PH-93/07,
hep-th/930317}
\lref\ELLIS{ E. Cremmer, S. Ferrara, C. Kounnas and D.V. Nanopoulos,
Phys. Lett. B133 (1983) 61; J. Ellis, A.B. Lahanas, D.V. Nanopoulos and
K. Tamvakis, Phys. Lett. B134 (1984) 429; J. Ellis, C. Kounnas and
D.V. Nanopoulos Nucl. Phys. B241 (1984) 406; B247 (1984) 373}
\lref\CLMO{ G. Lopes--Cardoso, D. L\"ust and T. Mohaupt, Berlin preprint
HUB--IEP--94/6, hep--th/9405002; I. Antoniadis, E. Gava, K. Narain and
T. Taylor, CPTH--A282.0194 (1994)}

\newsec{Introduction}

The  particle spectrum in supersymmetric versions of the
standard model (SM) is in general determined by soft
supersymmetry(SUSY)-breaking mass parameters like gaugino, squark, and slepton
masses. The possible numerical values of these soft terms are
only constrained either by experimental bounds on SUSY-masses or
some indirect theoretical arguments. The simplest supersymmetric
model, the minimal supersymmetric standard model (MSSM) assumes
certain universality of soft terms. SUSY-breaking  in the MSSM
is parametrized just by four parameters: a universal gaugino mass
$M$, a universal scalar mass $m$, a trilinear scalar
parameter $A$ and an extra bilinear scalar parameter $B$.
It is not clear what  SUSY-breaking dynamics could
be underlying the above simple choice of soft parameters but
one can construct SUSY-GUT type of theories in which the
above can be justified. On the other hand, the simplicity of
the MSSM choice is not compulsory. Many embeddings of the
supersymmetric standard model into bigger unified theories
lead to a more complicated soft term structure. That is the case \IL ,
e.g., of string models.

If low-energy supersymmetry is correct, eventually (and hopefully)
the spectrum of SUSY-particles will be measured and the structure of
SUSY-breaking soft terms will be subject to experimental test.
Particularly, the simple universal structure of the MSSM will be
tested and possible departures could be looked for. Apart from these
possible departures, one would learn what are the relative sizes of
$M,m,A,B...$. Imagine, e.g., that one finds that
$M\ll  m$. What information on the underlying theory could
one extract from that? Very little, unless we have a theory of SUSY-breaking
soft terms.

One can, of course, think of many possible models leading to
SUSY-breaking in a "hidden sector" of the theory leading in turn to
different results for the mentioned soft terms. Thus a
model-building avenue to face this problem does not look
very promising. We will assume instead that the {\it underlying
theory is a four-dimensional string} of some type and we will
try to make use of the general structure of the low-energy
sector of large classes of four-dimensional strings to
constrain the SUSY-breaking sector of the theory. We believe
this is a reasonable choice since heterotic strings remain the
only finite theories which can unify all the known interactions
including gravity.

The origin of supersymmetry breaking in strings is by itself an
outstanding problem. We know that such a process has to have a
non-perturbative origin since it is well known that SUSY is
preserved order by order in perturbation theory. On the other hand
very little is known about non-perturbative effects in
string theory, particularly in the four-dimensional case.
Thus one would say that strings do not look particularly
promising in trying to get information about the
SUSY-breaking sector of the theory.

The above view is unnecessarily pessimistic, as emphasized in \IL .
Indeed, if you think it over, the situation concerning
$SU(2)_L\times U(1)$ breaking in the SM is not in principle that different.
We do not really know for sure how the gauge symmetry of the SM is
broken. We just parametrize our ignorance by using a Higgs field
(either composite or elementary) with a non-vanishing vacuum
expectation value (VEV). the key ingredient here is knowing the
degrees of freedom (an $SU(2)_L$ doublet) involved in the process of
symmetry breaking. Once we know that, we can obtain all the
experimentally confirmed predictions of the SM.
To some extent, we face a similar
situation in the case of SUSY-breaking. We know the symmetry
($N=1$ supergravity) and what we need is to {\it identify the
degrees of freedom involved in the process of SUSY-breaking}.
The effect of SUSY-breaking will be then parametrized by the
VEVs of the auxiliary fields of the degrees of freedom identified.
Whereas in a non-string model we do not have the slightest idea
of what fields could be involved in SUSY-breaking, four-dimensional
{\it strings automatically have natural candidates for that job}:
the dilaton ($S$) and the moduli ($T_m$) chiral superfields
generically present in large classes of four-dimensional
supersymmetric heterotic strings. While other extra chiral multiplets
could also play a role in specific models, the dilaton and moduli
constitute in some way the {\it minimal possible SUSY-breaking sector
in string models}. Starting with this minimal sector one can also
study the possible role on SUSY-breaking of other extra chiral fields
(see the example discussed in section 8).

The above philosophy was first applied in ref. \IL ,
in which the idea of parametrizing the soft masses in terms of unknown
VEVs of the $S$ and $T_m$ auxiliary fields was already used.
Slightly less general formulae obtained in the context of the
gaugino condensation scenario \gcon \ were previously calculated in
refs.\FILQ ,\CFILQ. In ref. \CCM\ formulae for the soft
parameters including the one-loop corrected  K\"ahler potential
in the calculation were obtained and applied in particular to the gaugino
condensation scenario. The philosophy of \IL\ has also been
retaken by the authors of \KL\ who impose the cancellation of
the cosmological constant as an additional constraint and concentrate
mostly in the dilaton-dominated limit.
The purpose of the present paper is to
further develope the above approach for addressing the problem, trying
to provide a theory of soft terms which could enable us to interpret
the (future) experimental results on supersymmetric spectra.

The structure of the paper is as follows. In section 2 we discuss some
general formulae for the computation of soft terms in string models.
It turns out to be specially useful the introduction of a
"goldstino angle" whose value tells us where the dominant source of
SUSY-breaking resides. All formulae for soft parameters take
particularly simple forms when written in terms of this variable.
We also allow for a non-vanishing vacuum energy
and arbitrary complex phases in the relevant VEVs.
Allowing for these turn out to be important for some relevant
issues concerning soft terms. We also discuss the more
model dependent $B$ soft term for various solutions proposed
to generate a "$\mu $-term" in the low-energy theory. In section 3 we
compute soft terms for some large classes of models
including the large radius limit of Calabi-Yau (CY) \CHSW \ type of
compactifications and orbifold \DHVW\ type of models.  In section 4
the complex phases mentioned above play an important role. They may lead
to CP-violating effects  like an electric dipole moment for the
neutron (EDMN). Constraints on the soft terms from present experimental
bounds are also discussed. Some comments concerning flavour
changing neutral current (FCNC) effects in this type of models are
given in section 5.  In section 6 we discuss how the treatment one
gives to the cosmological constant problem has an important bearing on
the formulae obtained for the soft terms. In section 7 we consider the
low-energy renormalization group running of soft terms, the
low-energy sparticle spectra and the appropriate radiative
$SU(2)_L\times U(1)$ breaking. We do this for three scenarios which
intend to cover some generic situations in string models. They lead to
specific patterns for the SUSY-spectra which could be tested
in future colliders. The results one obtains for soft terms in the
present analysis of string models turn out to be in general different
to those found previously in explicit gaugino condensation scenarios
\CCM . The origin of this discrepancy is discussed in section 8. There
we also discuss in a simplified manner the modifications required
by our formalism if extra chiral fields beyond the
dilaton and the moduli are involved in breaking SUSY. Some final
comments and conclusions are left for section 9.

\newsec{ Structure of soft terms in generic string models}

\subsec{ General characteristics}

We will consider three types of massless chiral fields in
4-D string models. First there is the dilaton complex field
$S$, which couples universally in all models. The VEV of the real
part of S gives the inverse square of the tree level
gauge coupling constant. Then there are the moduli
fields, which we will denote by $T_m$.  Their VEVs parametrize
the size and shape of the compactifying variety of
the model. In particular, there is always a modulus field $T$ whose real
part gives the overall volume of the variety.
The main characteristic of the dilaton and the moduli is
that they have flat scalar potential (they correspond to marginal
deformations in the associated conformal field theory (CFT)). Moreover,
the fields $S$ and $T_m$ are singlets with respect to the gauge interactions of
interest, like the SM gauge interactions \footnote*{The $T_m$ fields may be
charged with respect to some extra $U(1)s$.
However, none of those $U(1)$s may be identified with
hypercharge.}. Finally there are the
charged fields, which we will denote by $C_i$. These should include the
quark, lepton and Higgs multiplets, and the corresponding scalar fields
will typically (although not always) have non-flat scalar potentials.

In order to proceed we need to make some simplifying assumptions. Some
of them are justified for generic 4-D strings. We will comment below what
changes are to be expected if the assumptions are relaxed. The first two
are the following:

1) Amongst the moduli $T_m$ we will concentrate on the overall modulus
$T$ whose classical value gives the size of the manifold. Apart from
simplicity, this modulus is the only one which is always necessarily present
in any $(0,2)$ (but left-right symmetric) 4-D strings.
 We believe that
studying the one modulus case is enough to get a feeling of the
most important physics of soft terms.
 We will disregard for the moment any mixing between the $S$ and $T$ fields
kinetic terms. In fact this is strictly correct in all 4-D strings at
tree level. However, it is known that this type of mixing may arise at one
loop level in some cases\IN,\DFKZ. On the other hand, these are loop effects
which should
be small and in fact can be easily incorporated in the analysis in some simple
cases (orbifolds) as shown  below.

Under the  above conditions, one can write for the K\"ahler potential
(to first order in the charged fields $C_i$) the following general
expression\W ,\FKP ,\DKL:
\eqn\KKK{
K(S,S^*,T,T^*,C_i,C_i^*)\ =\ -log(S+S^*)\ +\ K_0(T,T^*)\ +\
{\tilde K }^i_j (T,T^*)C_i {C^*}^j }
where the indices
$i,j$ label the charged matter fields.
For phenomenological reasons related to the
absence of flavour changing neutral currents (FCNC) in the effective
low-energy theory, from now on we will assume (see section 5 for a
discussion of this point) a diagonal
form for
 the part of the K\"ahler potential associated with the matter fields,
${\tilde K }^i_j = {\tilde K }_i \delta^i_j$.
The scalar potential  has the    following general form\CFGP
\eqn\pot{
V=e^{G} (G_{\alpha} (G^{-1})^{\alpha}_{\beta} G^{\beta}-3)
}
where $G=K+log|W|^2$ and $W$ stands for the complete superpotential
(including possibly non-perturbative effects) of the theory.

2) Only the generic marginal fields of the theory
($S$ and $T$) contribute to SUSY-breaking through non-vanishing
VEVs for their auxiliary fields. Since these fields  (unlike the generic
charged fields) have vanishing perturbative potential it is reasonable to
expect them to play an important role in supersymmetry-breaking.
We are aware that  there could be other marginal deformations with
vanishing scalar potential (e.g., continuous Wilson lines) in some
specific models. The modifications needed in our formulation in that case
are schematically discussed in section 8.

Under the above condition, the vacuum expectation value of the scalar
potential (i.e. the cosmological constant) is
\eqn\VVV{
\eqalign{
V_0\ & =\ e^G  (G_S^S)^{-1}\ |G^S|^2\ +\ e^G (G_T^T)^{-1}\ |G^T|^2
\ -\ 3\ e^G \cr
& =\ G_S^S |F^S|^2\ +\ G_T^T |F^T|^2 \ -\ 3 e^G \cr
}
}
where $e^G = m_{3/2}^2$ is the gravitino mass-squared. Of course,
the first two terms in the right hand side of \VVV\ represent the
contributions of the $S$ and $T$ auxiliary fields,
$F^S=e^{G/2}(G^S_S)^{-1} G^S$ and
$F^T=e^{G/2}(G^T_T)^{-1} G^T$.

As we will show below, it is important to know what field, either $S$ or $T$,
plays the predominant role in the process of SUSY-breaking. This
will have relevant consequences in determining
the pattern of soft terms, and therefore the spectrum of physical
particles. That is why
it is very useful to  define an angle $\theta $
in the following way (consistently with eq.\VVV):
\eqn\ans{
\eqalign{
(G^S_S)^{1/2}\ F^S\ & =\ {\sqrt 3}C m_{3/2}\ e^{i\alpha _S}sin\theta  \cr
(G^T_T)^{1/2}\ F^T\ & =\ {\sqrt 3}C m_{3/2}\ e^{i\alpha _T}cos\theta \cr
}
}
where $\alpha_S,\alpha_T$ are the phases of $F^S$ and  $F^T$, and the
constant $C$ is defined as follows:
\eqn\ccc{
C^2\ =\ 1\ +\ {{V_0}\over {3m_{3/2}^2}}
.}
If the cosmological constant $V_0$ is assumed to vanish, one has $C=1$, but we
prefer for the moment to leave it undetermined. We will show later on that, in
specific analyses, the way one deals with the
cosmological constant problem is important.

Notice that, with the above assumptions, the goldstino field which is swallowed
by the gravitino in the process of supersymmetry breaking is proportional
to
\eqn\gold{
{\tilde {\eta }}\ =\ sin\theta \ {\tilde S}\ +\ cos\theta \ {\tilde T}
}
where ${\tilde S}$ and ${\tilde T}$ are the canonically normalized
fermionic partners of the
scalar fields $S$ and $T$ (we have reabsorbed here the phases by redefinitions
of the fermions ${\tilde S},{\tilde T}$). Thus the angle defined above
may be appropriately termed {\it goldstino angle} and has a clear
physical interpretation as a mixing angle.

\subsec{Computation of soft terms}

As it is well known, spontaneous SUSY-breaking in the supergravity theory
leads to an effective low-energy theory with global supersymmetry
explicitly broken by a set of soft terms. We also recall that
the passage to the effective low-energy theory involves a number
of rescalings.
In particular, canonically normalized charged fields are
given by $ {\hat C}_i = {\tilde K }_i^{1/2} C_i$ and
the original superpotential associated with those fields,  $W(C_i)$,
will give rise to an effective low-energy superpotential
${\hat W}({\hat C}_i)={{W^*} \over {|W|} }e^{K/2} W(C_i)$.
We will now compute explicitely
the SUSY-breaking soft terms as functions of $m_{3/2},C$ and $\theta$.
In particular, bosonic soft terms are obtained by
expanding the scalar potential \pot, using the form \KKK\ of
the  K\"ahler potential.

{\it  a) Scalar masses}

One type of soft terms which arise from the expansion of $V$
are mass terms of the form $m_i^2 |{\hat C}_i|^2$.
Soft scalar masses for models with ${\tilde K}_i\propto (T+T^*)^{n_i}$ were
first computed in refs.\CFILQ,\IL,\CCM. In ref.\KL\ the equivalent
results for a generic ${\tilde K}_i^j$ were obtained in terms of the K\"ahler
geometry of the supergravity action (see also \SW ). Using our assumptions
above, the expression for the masses takes the form

\eqn\msof{
m_i^2\ =\ 2m_{3/2}^2\ (C^2-1) \ +\ m_{3/2}^2C^2(1 +N_i(T,T^*)cos^2\theta )
}
where $N_i(T,T^*)$ is defined as \IL
\eqn\NNN{
N_i(T,T^*)\  =\ {3 \over {{K_0}_T^T} }
( { { {{\tilde K }_i}_T {{\tilde K }_i}^T }
\over { {{\tilde K }_i}^2 }  }  \ -\
{ {{{\tilde K}_i}^T_T}\over  {{\tilde K }_i}  } )
\ = \
-3{{ (log {\tilde K}_i)^T_T} \over {{K_0}_T^T} }
}
This formula for $N_i(T,T^*)$ looks complicated but it becomes very simple
in some classes of string models and/or in some limits. In particular,
$N_i$ is related to the curvature of the K\"ahler manifold parametrized by the
above K\"ahler potential. For manifolds of constant curvature (like in
the orbifold case) the $N_i$ are constants, independent of $T$. More precisely,
they correspond to the modular weights of the charged fields, which are
normally negative integer numbers (see ref.\IL\ for a classification of
possible modular weights of charged fields in orbifolds). In more complicated
four-dimensional strings like those based on Calabi-Yau manifolds, the
$N_i(T,T^*)$ functions are complicated expressions in which world-sheet
instanton effects play an important role. In the case of $(2,2)$ Calabi-Yau
manifolds, for the large $T$ limit it turns out that
$N_i(T,T^*)\rightarrow -1$\CD,\C. We will come back to the evaluation of
the $N_i$ in specific string models later on.

{\it b)  The A term}

Let us assume that the charged fields have trilinear superpotential terms
of the form $W_{ijk}=h_{ijk}(T)C_iC_jC_k$. Then the effective
low-energy theory will contain superpotential terms
of the same form, ${\hat W}_{ijk}={\hat h}_{ijk}(T) {\hat C}_i{\hat C}_j{\hat
C}
_k$,
where $ {\hat h}_{ijk}=  h_{ijk}{{W^*} \over {|W|} }
e^{K/2}({\tilde K }_i{\tilde K }_j{\tilde K }_k)^{-1/2}$.
Also, the expansion of the potential $V$ will give rise
to trilinear SUSY-breaking terms of the form $ A_{ijk}\ {\hat h}_{ijk}
({\hat C}_i{\hat C}_j{\hat C}_k)\ +\ h.c.$, where $ A_{ijk}$ is a dimensionful
parameter of order the gravitino mass.  One finds

\eqn\AAF{
A_{ijk}\ =\ -{\sqrt 3}m_{3/2}C(e^{-i\alpha _S}sin\theta \  +
\ e^{-i\alpha _T} \omega_{ijk}(T,T^*) cos\theta) }
where
\eqn\omo{
{\omega _{ijk}}\ =\ ({K_0}_T^T)^{-1/2}
( {\sum _{l=i,j,k}} {{{\tilde K }_l^T}\over {{\tilde K }_l}}
\ -\ K_0^T\ -\ {{h_{ijk}^T}\over {h_{ijk}}} \ )
}
Notice that now the phases $\alpha _S$ and $\alpha _T$ may become relevant.
As in the case of the scalar masses, the above expression becomes much simpler
in specific four-dimensional strings and/or in the large $T$ limit. We will
see later on examples of these. The $A$-term for models with ${\tilde K}_i
\propto
  (T+T^*)^{n_i}$ were
first obtained in refs.\CFILQ,\CCM.

{\it c)  Gaugino masses}

Physical gaugino masses $M_a$ for the canonically normalized gaugino
fields are given by $M_a= {1\over 2}(Re f_a)^{-1}
 e^{G/2}f_a^{\alpha} (G^{-1})_{\alpha}^{\beta} G_{\beta}$.
We will assume that the $N=1$
gauge kinetic functions $f_a$ only depend on the $S$ and $T$ chiral
fields (in fact it is enough to assume that other possible chiral fields do
not contribute to SUSY-breaking). Under these circumstances one gets
\eqn\MMM{
\eqalign{
M_a\ &=\  {1\over {2 Ref_a}} e^{G\over 2}( f_a^S(G_S^S)^{-1}\ G_S\ +\
 f_a^T(G_T^T)^{-1}
\ G_T )\ = \cr
&=\ {{C{\sqrt 3}}\over {2 Ref_a}} m_{3/2}
( f_a^S(G_S^S)^{-1/2}\ e^{-i\alpha _S}sin\theta \
 + \ f_a^T(G_T^T)^{-1/2}\ e^{-i\alpha _T}cos\theta ).\cr }
}
The tree-level expression for $f_a$ for any four-dimensional string is well
known,
$f_a=k_aS$, where $k_a$ is the Kac-Moody level of the gauge factor (in the
phenomenological analysis of section 7 we will restrict the discussion
to the level one case $k_3=k_2={3 \over 5} \ k_1 =1$ for the MSSM).
A possible $T$ dependence may appear at one-loop \IN .
Then, the one-loop corrected $f_a$ function may be parametrized by
\eqn\fff{
f_a\ =\ k_aS\ -\ {1\over {16\pi ^2}}B'_a log(\Delta (T))^2 \ \ .
}
In choosing this parametrization we have been inspired by explicit
computations in the orbifold case \DKLD , as we will
discuss later on. Here $B'_a$ is a numerical coefficient and $\Delta $ contains
all the dependence on $T$ ("threshold correction"). Using this general
expression one finds for the gaugino masses the result
\eqn\gaug{
M_a\ =\  {{C{\sqrt 3}}\over {Ref_a}} \  m_{3/2}
 (k_a ReS e^{-i\alpha _S}sin\theta
 \ -\ {{B'_a}\over {16\pi ^2}}  {1 \over {\Delta}}
{{\partial \Delta} \over {\partial T}}
({K_0}_T^T)^{-1/2}e^{-i\alpha _T}cos\theta \ )
}
where $Ref_a$ are the inverse squared gauge coupling constants
at the string scale.
 (See refs.\FILQ ,\IL , \CCM and \KL \ for previous computations of soft
gaugino masses in string models).

Notice that, differently from the present case, we did not
include one-loop effects in the computation of the other
soft terms in this section. The motivation is the following:
one-loop corrections to the
$f_a$ function are known explicitely in some four-dimensional strings,
and moreover higher-loop corrections are vanishing, whereas
computing the one-loop-corrected bosonic soft terms would
require knowledge of the one-loop-corrected K\"ahler
potential, whose form is not available in the general case.
However, for some orbifolds models, well motivated conjectures give the
form of the one-loop-corrected K\"ahler potential: in particular,
 mixing appears between the S and T kinetic terms. We
will take into account such corrections in the next section.
They are normally negligible, but may be
important for small $sin\theta $, as we will see in specific cases.

{\it d)  The B-term}

Apart from the $m_i$, $A_{ijk}$ and $M_a$ soft parameters, soft bosonic
bilinear terms can be present. In the case of the MSSM, in particular,
all the symmetries of the low-energy theory allow for a superpotential
term  coupling the two Higgs doublets of the form
\eqn\mum{
{\hat W}_{\mu} = {\hat\mu} {\hat H} {\hat{\bar H}}
}
The associated soft-breaking term in the scalar potential will have the form
\eqn\BBB{
B {\hat\mu} \ {\hat H} {\hat{\bar H}}\ +\ h.c.
}
where $B$ is a dimensionful parameter of order the gravitino mass.
It is well known that, in order to get appropriate
$SU(2)_L\times U(1)$ breaking, the $\mu $ parameter has to be of the same order
of
magnitude ($m_{3/2}$) as the SUSY-breaking soft terms discussed above.
This is in general
unexpected since the $\mu $ term is a supersymmetric term whereas the other
soft terms are originated after SUSY-breaking. This is sometimes called
"the $\mu $ problem", the reason why $\mu$ should be of order the soft terms.
We will mention here the
three main scenarios considered up to now in order to solve this
problem \footnote*{Another possible solution via SUSY breaking in
string perturbation theory can be found in ref.\AMQ.}:

i) Adding an extra singlet $N$ with superpotential couplings of the form
$W_N=\lambda NH{\bar H}+\sigma N^3$. Then, the role of $\mu $ is played
by $\lambda <N>$ and the role of $B$ is played by $A_{\lambda }$. The
resulting model is no longer the MSSM but the simplest extension leading
to similar physical results (except for the neutralino and Higgs sector)\NSW.

ii) Explicit non-perturbative SUSY-breaking mechanisms like gaugino
condensation may naturally induce effective $\mu $ parameters of the order
of the gravitino mass\KN,\MUCM.
In this case the supergravity superpotential already contains a term
\eqn\WMU{
W_{\mu} = \mu(S,T) H {\bar H}
}
The low-energy theory will have a corresponding superpotential term
as in eq.\mum, where
${\hat \mu}$ is related to $\mu$ via a rescaling,  ${\hat \mu}=
{{W^*} \over {|W|} }{\mu} e^{K/2}({\tilde K }_H {\tilde K }_{\bar H})^{-1/2}$,
analogously to the trilinear coupling case. We will call $B_{\mu}$
the associated $B$ parameter.

iii) If certain additional terms are present in the tree level K\"ahler
potential \KKK , an effective low-energy $\mu $ term may naturally be
generated of the order of magnitude of the gravitino mass. In particular,
if an additional term
\eqn\giu{
\delta K\ =\ Z(T,T^*)H{\bar H}\  +\ h.c.
}
is present in $K$, an effective  $\mu $-term is generated in the
low-energy theory\GM,\MUCM. Here it can be described again by a superpotential
term of the form \mum, where now ${\hat \mu}=(W Z)
{{W^*} \over {|W|} } e^{K/2}({\tilde K }_H {\tilde K }_{\bar H})^{-1/2}$.
We will call $B_Z$ the associated $B$ parameter.

In the simple case $Z=constant$ it is trivial to see the equivalence with
the previous mechanism\MUCM. Indeed, in the case now considered, the
 supergravity
theory is equivalent to one with K\"ahler potential
$K$ (without the additional term $\delta K$ above) and
superpotential $W e^{Z H{\bar H}}$, since the function
$G=K+log|W|^2$ is the same for both. After expanding the exponential,
the superpotential will have a contribution
\eqn\WZZ{
W_Z= (W Z) \ H {\bar H}
}
i.e. a term of the type \WMU\ with $\mu(S,T)=W(S,T) Z$. Notice
also that a constant $Z$ is equivalent to an $S$ and $T$ dependent $\mu$.

Contributions of the type \giu\ do not seem to be present
in $(2,2)$ orbifold-type
compactifications\KL, but they are in general present in the non-singular
Calabi-Yau case.

We will not say anything about the first alternative which introduces a
new singlet to solve the problem. In this case there are no bilinear couplings
and hence the only soft terms present are the ones already described above.
On the other hand, one can compute the form of the $B$ parameter for the
other two cases. The results are
\eqn\bmu{
\eqalign{
B_{\mu}\ &=m_{3/2}(\ -1\ -C{\sqrt 3}e^{-i\alpha _S}sin\theta
(1-{{\mu ^S}\over {\mu }}
(S+S^*))\ +\cr
\ &C{\sqrt 3}e^{-i\alpha _T}cos\theta
({K_0}_T^T)^{-1/2}( K_0^T +\ {{\mu ^T}\over {\mu }}
-{{{\tilde K }_H^T}\over {{\tilde K }_H}}
-{{{\tilde K }_{\bar H}^T}\over {{\tilde K }_{\bar H}}}
 )) \cr }
}
\eqn\gii{
\eqalign{
B_Z\ &= m_{ 3/2} ({ 3 \over K}(C^2-1) +
\ 2\ +C {\sqrt 3}e^{-i\alpha _T}cos\theta
({K_0}_T^T)^{-1/2}({{Z^T}\over Z}-{{{\tilde K }_H^T}\over
{{\tilde K }_H}}-{{{\tilde K }_{\bar H}^T}\over {{\tilde K }_{\bar H}}})\ -\
 \cr
&C{\sqrt 3}e^{i\alpha _T}cos\theta ({K_0}_T^T)^{-1/2}{{Z_T}\over Z}
\ +\ C^2 3({K_0}_T^T)^{-1}
cos^2\theta ({{Z_T}\over Z}({{{\tilde K }_H^T}\over
{{\tilde K }_H}}+{{{\tilde K }_{\bar H}^T}\over {{\tilde K }_{\bar H}}})\
-\ {{Z^T_T}\over Z})) \cr }
}
In both eq.\bmu\ and \gii\  there are several unknown functions
and hence these expressions have  limited predictivity unless
one considers some particular limits. Let us make here some simplifying
assumptions in order to get somewhat simpler formulae. Let us  assume that the
dependence of the function $\mu (S,T)$ on $S$ and $T$ is weak close to the
minimum of the scalar potential so that one can set
the ratios $\mu ^S/\mu =\mu ^T /\mu =0$.
One then finds that eq.\bmu\ is simplified to
\eqn\bbm{
B_{\mu }\ = m_{ 3/2}(\ -1\ -\ C{\sqrt 3}e^{-i\alpha _S}sin\theta
 \ -\ C{\sqrt 3}e^{-i\alpha _T}cos\theta  ({K_0}_T^T)^{-1/2}
({{{\tilde K }_H^T}\over {{\tilde K }_H}}+
{{{\tilde K }_{\bar H}^T}\over {{\tilde K }_{\bar H}}}-K_0^T))
 }
This expression should only be taken as indicative of a possible
$\theta$--dependence of the $B$--parameter since indeed there is no
obvious reason why one should have $\mu ^S/\mu \simeq \mu ^T /\mu \simeq 0$.
On the other hand, in the case of the third mechanism with an extra term
in the K\"ahler potential, the expression for $B_Z$ simplifies a lot in
the large $T$ limit of Calabi--Yau strings, as we will see in the next
section. Notice that it is
%
%
conceivable that both mechanisms could be present simultaneously for
some string models. In that case the general expression is more complicated
(it is $not$ just the sum $B_{\mu }+B_Z$) and can be easily obtained
from the above formulae.
We will confine ourselves to considering the two
possibilities separately. We expect that, if both mechanisms were
simultaneously present, the physical results would be somewhere in-between
both extremes.

As a final comment, notice that
the results for the $B$ parameter obtained for $\mu =
const.$ (in approach ii)) and $Z =const.$  (in approach iii))
do not coincide  (see eqs.\bbm, and \gii\ with $Z^T=Z_T=0$).
The reason is that a constant $Z$ is not equivalent to
a constant $\mu$, as we remarked after eq.\WZZ.

\subsec{The $sin\theta =1$ (dilaton-dominated) limit}

Before going into specific classes of string models, it is worth studying the
interesting limit $sin\theta =1$, corresponding to the case where
the dilaton sector is the source of all the SUSY-breaking (see
eq.\ans). Since the
dilaton couples in a universal manner to all particles, this limit is
quite model independent.
The existence of a universal limit was remarked in ref.\IL. In ref.\KL\
it was explicitly remarked that this universal limit corresponds to
dilaton--dominance and specific formulae for this limit were obtained by
imposing
the additional constraint of cancellation
of the cosmological constant (i.e., $C=1$).
Using eqs.\gaug,\msof,\AAF\
one finds the following simple expressions for the
soft terms:
\eqn\mag{
\eqalign{
M_a\ & =\ {\sqrt 3}Cm_{3/2}{{k_a ReS}\over {Ref_a}}e^{-i\alpha _S} \cr
m_i^2\ & =\ C^2 m_{3/2}^2\ +\ 2m_{3/2}^2(C^2-1) \cr
A_{ijk}\ & =\ -{\sqrt 3}Cm_{3/2}e^{-i\alpha _S} \cr }
}
Similar expressions for the $C=1$, $\alpha_S$=0 case were obtained in
ref.\KL\ except for a factor 1/2 mistakenly given for the gaugino masses.
Notice that the scalar masses and the $A$ terms are universal,
whereas the gaugino masses may be slightly non-universal
since non-negligible threshold effects might be present.
Finally, the situation for the $B$-term is much more uncertain,
as discussed above. If one takes the formulae \bbm\
and \gii\ as      indicative,  one finds
\eqn\bch{
B_{\mu }\ =\ m_{ 3/2}( -1\ -\ {\sqrt 3}Ce^{-i\alpha _S})\
=\ A\ -\ m_{3/2}
}
\eqn\bzc{
B_Z\ =\ m_{ 3/2}( 2 + { 3 \over K}(C^2-1) \ ) \  .
}
Notice that the expression for $B_{\mu }$ obtained in this limit coincides with
the one obtained in supergravity models with canonical kinetic terms for
the matter fields\HLW  .
 It is obvious that this
limit is quite predictive \KL. For a vanishing cosmological constant (i.e.
$C=1$), the soft terms are in the ratio $m_i:M_a:A=1:{\sqrt 3}:{-\sqrt 3}$
up to small threshold effect corrections (and neglecting phases). This will
result in definite patterns for the low-energy particle spectra, as we will see
in Section 7.

 \newsec{  Computing Soft Terms in Specific String Models}

In order to obtain more concrete expressions for the soft terms one has to
compute the functions $N_i(T,T^*)$, $\omega _{ijk}(T)$ and $\Delta (T)$. In
order
to evaluate these functions one needs a minimum of information about the
K\"ahler potential $K$, the structure of Yukawa couplings $h_{ijk}(T)$ and the
one-loop threshold corrections $\Delta (T)$. This type of information is only
known for some classes of four-dimensional strings which deserve special
attention. We thus will concentrate here in two large classes of models:
large $T$ limit of Calabi-Yau compactifications\CHSW \ and orbifold
compactifications\DHVW.
We will describe the general pattern of soft terms in these large classes of
string models in turn.

\subsec{Large $T$ limit of Calabi-Yau compactifications}

Little is known about the general form of the K\"ahler potential and couplings
of generic Calabi-Yau $(2,2)$ compactifications. Only a few examples
(most notably, the quintic in $CP^4$) have been worked out \PX \ in some detail
\CD \
and show formidable complexity due to the world-sheet instanton contributions
to the K\"ahler potential.
On the other hand, a few generic facts concerning these models are known
for the large $T$ limit\CD,\C. Large $T$, in practice, does not really mean
$T\rightarrow \infty$, since the world--sheet instanton corrections are
exponentially suppressed. For values $|T|\geq 2-3$ these world--sheet instanton
contributions can often be neglected and, in this sense these $T$--values are
already large. It is true that $|T|$ cannot be infinitely large, since
otherwise the quantum corrections to the gauge coupling constants (string
threshold corrections) may be too large and spoil perturbation theory. The
maximum allowed $|T|$ not spoiling perturbation theory is something which
is model dependent but is expected to be much bigger than one since, after
all, the threshold corrections are loop effects. In explicit orbifold
examples it was found in refs.\ILR,\IL\ that $|T|\leq 20-30$ is enough
to remain in
the perturbative regime. When we talk about the large $T$ limit in what
follows we will thus assume $T$--values which do not spoil perturbativeness.
In this limit the K\"ahler potential $K$ gets a
particularly simple form:
\eqn\cyt{
K(T\rightarrow \infty )\ =\ -log(S+S^*)\ -\ 3log(T+T^*)\ +\
\sum _{i} |C_i|^2 (T+T^*)^{-1}\ \ .
}
This leads to a number of important simplifications. Plugging the above
expression
into eqs.\NNN\  and \omo\  one finds
\eqn\NCY{
N_i(T,T^*)\longrightarrow  \ -1
}
\eqn\ocy{
\omega _{ijk}(T, T^*)\longrightarrow \ {{-(T+T^*)}\over {\sqrt 3}}{{h_{ijk}^T}
\over {h_{ijk}}}
}
It can be further argued that, in the large $T$ limit, the non-vanishing Yukawa
couplings tend (exponentially) to constants, as computed in specific examples.
Then one can
take $\omega _{ijk}\rightarrow 0$ in the mentioned limit.
Using the above information in eqs.\msof, \AAF, \gaug \
one gets the following $T\rightarrow \infty $ formulae
\eqn\socy{
\eqalign{
m_i^2\ &=\ m_{3/2}^2C^2sin^2\theta\ +\ 2m_{3/2}^2(C^2-1)  \cr
M_a\ &=\ {\sqrt 3}Cm_{3/2}{{k_a ReS}\over {Ref_a}}e^{-i\alpha _S}sin\theta \cr
A_{ijk}\ &=\ -{\sqrt 3}Cm_{3/2}e^{-i\alpha _S}sin\theta \cr }
}
where we have ignored the possible one-loop corrections to these
formulae.
It is interesting to remark that in this large $T$ limit of CY-type
compactifications
the results obtained for the soft terms are quite similar to those  in
eqs.\mag\ obtained in a model-independent manner for $sin\theta =1$. The
role of dimensionful parameter is played now by $m_{3/2}sin\theta $ (for $C=1$)
instead of simply $m_{3/2}$.
Thus {\it dilaton-dominated SUSY-breaking is not the only situation in which
universal soft scalar masses are obtained}, as the present model exemplifies.
Anyhow we point out that in the case with several moduli the situation is
much more cumbersome and one is forced to define new goldstino angles.
If we allow generic values for these angles we obtain a deviation from
the previous universal behaviour. It is also interesting
to notice how, for $C=1$,  all these terms
tend to zero at the same speed as $sin\theta \rightarrow 0$, even for a finite
value of $m_{3/2}$. However, for $\sin \theta =0$ the one-loop corrections to
the K\"ahler potential cannot probably be neglected, and care should be taken
before
getting any definite conclusion (see the case of one-loop orbifold corrections
discussed below).

Concerning the $B$-parameter, great simplifications do also occur. For example,
formula \bbm \ now reads

\eqn\bcy{
B_{\mu }\ =m_{3/2}(\ -1\ -\ C{\sqrt 3} e^{-i\alpha _S}sin\theta \ -
C e^{-i\alpha _T} cos\theta)
}
%
%
%
The equivalent formula for the alternative mechanism in which there is an
extra term in the K\"ahler potential originating a $\mu$--term is also
very much simplified. Indeed, in the case of a (2,2) Calabi--Yau string,
the function $Z(T,T^*)$
can be related to some of the Yukawa couplings of the theory
by Ward identities \KL . Then for large $(T+T^*)$ one expects\footnote*{We
thank the referee for pointing out an error in this comment in the preprint.
Similar $T$--dependence of $Z$ has also been recently found in some orbifold
models \CLMO}
$Z(T,T^*)\longrightarrow  \ (T+T^*)^{-1}$, and
formula \gii\ collapses to the simple result:
\eqn\apel{
B_Z\ = m_{ 3/2}({ 3 \over K}(C^2-1) +
\ 2\ +\ 2C cos\alpha _T cos\theta)
}

The above statements concerning the large $T$ limit of Calabi-Yau
compactifications
are known to be true for $(2,2)$ models, which yield a gauge group $E_6\times
E_
8$.
In order to make contact with the standard model one has to break this
structure with Wilson line gauge symmetry breaking and/or use $(0,2)$ type
compactifications. However, it is reasonable to expect that the general
structure
in eq.\cyt\ will still apply in these more complicated cases and, hence,
eqs.\socy\ will still hold.

Notice that the $sin\theta\rightarrow 0$ limit of the large $T$ Calabi--Yau
strings is different from the "no--scale" supergravity models discussed in
the literature \ELLIS. Although in both models (for $C$=1) one has at the
tree level $m_i=A=0$, the behaviour of the gaugino mass is totally different.
In the no--scale models the gaugino mass is non--vanishing and constitutes
the only source of supersymmetry--breaking whereas in the present class of
models the gaugino mass also vanishes.

\subsec{Orbifold compactifications}

 In the case of orbifold four-dimensional strings, a number of simplifications
occur without needing to go to the large $T$ limit. The K\"ahler potential
has now the general form (for small $|C_i|$)\W,\FKP,\DKL
\eqn\kor{
K(S,S^*,T,T^*,C_i,C_i^*)\ =\ -log(S+S^*)\ -3log(T+T^*)\ +\
\sum _i |C_i|^2 (T+T^*)^{n_i}
}
where the $n_i$ are integers, sometimes called modular weights \CFILQ ,\IL \
of the matter
fields.
It is important to remark that, unlike the case of smooth Calabi-Yau models,
the above T-dependence does not get corrections from world sheet instantons and
is equally valid for small and large $T$. In fact, general orbifold
models have a symmetry ("target-space duality") which relates small to large
$ReT$\FLT. This is a discrete infinite subgroup of $SL(2,R)$ in which
$T$ plays the role of modulus. For the overall field $T$ here considered,
the target-space duality group will be either
the modular group $SL(2,Z)$ or a subgroup of it
(in some cases, if quantized Wilson lines are present). Under $SL(2,Z)$
the modulus transforms like
\eqn\slz{
T\ \rightarrow \ {{aT-ib}\over {icT+d}} \ \ ;\ \ ad-bc\ =\ 1\ ,\ a,b,c,d\in
{\bf
 Z}
}
the dilaton $S$ field is invariant at tree-level and the matter fields
transform like
\eqn\mod{
C_i\ \rightarrow \ \ (icT+d)^{n_i}\ C_i
}
up to constant matrices which are not relevant for the present analysis.
Eq.\mod
\
explains why the integers $n_i$ are called modular weights. With the above
transformation properties the $G$ function is modular invariant (if the
superpotential $W$ has modular weight $-3$). The modular weights of the matter
fields
$n_i$ are normally negative integers (see ref. \IL ). For example, in the case
of $Z_N$ orbifolds the possible modular weights of matter fields are
$-1,-2,-3,-
4,-5$.
Fields belonging to the untwisted sector have $n_i=-1$.
Fields in twisted sectors of the orbifold but  without
oscillators have usually modular weight $=-2$ and those with oscillators have
$n_i\leq -3$. Notice that, with matter fields in the untwisted sector, the
resulting K\"ahler potential is analogous to the one obtained in the large $T$
limit of Calabi-Yau models.

Using eqs.\kor, \msof, \AAF \  one obtains for the soft scalar masses
and for the $A$ parameters:
\eqn\mor{
m_i^2\ =\ m_{3/2}^2C^2(1\ +\ n_icos^2\theta )\ +\ 2m_{3/2}^2(C^2-1)
}
\eqn\aor{
A_{ijk}\ =\ -{\sqrt 3} C m_{3/2}( e^{-i\alpha _S}sin\theta\ +\
 e^{-i\alpha _T} cos\theta\ \omega _{ijk}(T)\ )\ ,
}
where
\eqn\oor{
\omega _{ijk}(T)\ =\ {1\over {\sqrt 3}}(3+n_i+n_j+n_k\ -(T+T^*)
{{h_{ijk}^T}\over {h_{ijk}}}\ )\ .
}
Notice the explicit dependence of the $A$-parameters on the modular weights
of the particles appearing in the Yukawa coupling.
Moreover, the last term in eq.\oor\ drops for Yukawa couplings
involving either untwisted fields or (for large $T$) twisted fields associated
to the $same$
fixed point. The reason is that the Yukawa couplings
are constants or tend exponentially to constants, respectively\HV.
Since in the phenomenological analysis of section 7 the relevant
couplings will be of this type\footnote*{The $A$ term which is
relevant to electroweak symmetry breaking is the one associated
to the top-quark Yukawa coupling. If the fields are twisted
they should be associated to the same fixed point in order
to obtain the largest possible value of the coupling, otherwise
it would be exponentially suppressed\CM.}, we drop such term from now on.

The threshold correction function $\Delta (T)$ has been computed for
$(2,2)$ $Z_N$ and $Z_N\times Z_M$ orbifolds in refs.\K,\DKLD,\L,\DFKZ,\MS.
The result has the form in eq.\fff\   with
\eqn\for{
 \eqalign{
\Delta (T)\ &=\ \eta ^2(T)  \cr
B_a'\ &=\ b_a'\ -k_a\delta _{GS}\cr }
}
where $\eta (T)$ is the well known Dedekind function which admits the
representation
\eqn\ded{
\eta (T)\ =\ e^{-\pi T/12}\ \Pi _{n=1}^{\infty}(1-e^{2\pi nT})\ .
}
In eq.\for\ $k_a$ is the Kac-Moody level of the gauge group $G_a$, $\delta
_{GS}
$ is
a group independent (but model dependent) constant and
\eqn\bpr{
\eqalign{
b_a'\ &=\ -3C(G_a)\ +\ \sum _i T_a(C_i)(3 + 2n_i)  \cr
&=\ b_a\ +\ 2\sum _i T_a(C_i)(1 + n_i) \cr }\
}
where
$C(G_a)$ denotes the quadratic Casimir in the adjoint representation
of $G_a$,  $T_a(C_i)$ is defined by $Tr(T^{\alpha} T^{\beta} )
=T_a(C_i){\delta}^{\alpha \beta}$ ($T^{\alpha}=$ generators of $G_a$
in the $C_i$ representation)
and the sum runs over all the massless charged chiral fields.
Notice that $b_a$ is nothing but the $N=1$ one-loop $\beta $-function
coefficient
of the $G_a$ gauge coupling and that $b_a'=b_a$ when all matter fields have
modular weights $n_i=-1$ (as for untwisted states). While $b_a'$ may be
computed
in terms of the effective low-energy degrees of freedom of the theory,
$\delta _{GS}$ is a model dependent quantity (a negative integer in the case
of the overall modulus $T$). Its presence is associated to the cancellation of
the one-loop anomalies of the theory (we direct the reader to refs.\DFKZ,\LO\
for an explanation of these points which are not crucial for the understanding
of
what follows). Using eqs.\kor, \gaug\ and the above expressions
for $B_a'$ and $\Delta (T)$ one gets the result
\eqn\gor{
M_a\ =\ {\sqrt 3}m_{3/2}C({{k_a ReS}\over {Ref_a}}e^{-i\alpha _S}sin\theta \ +\
e^{-i\alpha _T}cos\theta ({{B_a'(T+T^*){\hat {G_2}}(T,T^*)}\over
{32\pi ^3{\sqrt 3}Ref_a}} ) \ )
}
where ${\hat {G_2}}$ is the non-holomorphic Eisenstein function which may be
defined by ${\hat {G_2}}=G_2(T)-2\pi/(T+T^*)$. Here $G_2$ is the holomorphic
Eisenstein form which is related to the Dedekind function by
$G_2(T)=-4\pi(\partial \eta (T)/\partial T)(\eta (T))^{-1}$.
In fact, using eq.\gaug\
one gets eq.\gor\ with $G_2$ instead of ${\hat {G_2}}$. However, one gets the
complete result in eq.\gor\ when one includes the one-loop contribution of
massless
fields (see ref.\IL).

The gauge threshold correction effects discussed above were in fact worked out
for the $(2,2)$ case although they are expected to be valid for more general
cases
(see e.g. ref.\IL\ for a general discussion on this point).

Concerning the $B$-parameter, for e.g. the second mechanism eq.\bbm \ reads
\eqn\bsi{
B_{\mu }\ = m_{ 3/2}(\ -1\ -\ C{\sqrt 3}e^{-i\alpha _S}sin\theta
 \ -\ C e^{-i\alpha _T}cos\theta  (3+n_H+n_{\bar H})) \ .
}

As discussed in the previous section,
there is a slight inconsistency in using one-loop formulae
for the gaugino masses whereas for the other
soft terms $m_i$, $A$ and $B$ we use only the tree level result. In fact this
is
 not that
important since normally the one-loop corrections are small for the other two
terms.
Anyway, they may be evaluated knowing the one-loop (string) corrections to the
K\"ahler potential. General arguments applicable to orbifolds allow us to write
the one-loop corrected K\"ahler potential for orbifolds by making the
replacement\DFKZ
\eqn\yyy{
S+S^*\ \longrightarrow \ \ Y\ =\ S+S^*\ -\
{{\delta _{GS}}\over {8\pi ^2}}log(T+T^*)\
}
in eq.\kor. As a consequence, mixing appears between the $S$ and $T$ kinetic
terms, and the expression \VVV\ for the VEV of the scalar
potential gets modified as follows:
\eqn\VMIX{
\eqalign{
V_0\ & =\ e^G  Y^2 |G^S|^2 +  e^G
{ {(T+T^*)^2} \over  3} (1- { {\delta _{GS}} \over {24\pi^2 Y} } )^{-1}
 |G^T+  { {\delta _{GS}} \over {8\pi^2(T+T^*)} } G^S|^2
\ -\ 3  e^G  \cr
 & =\ {1 \over {Y^2}}
 |F^S - { {\delta _{GS}} \over {8\pi^2(T+T^*)} } F^T|^2 +
{ 3 \over  {(T+T^*)^2} } (1- { {\delta _{GS}} \over {24\pi^2 Y} } ) |F^T|^2
\ -\ 3 e^G\
}
}
Analogously to eq.\VVV, we have written $V_0$ also in terms of the $S$ and $T$
auxiliary fields $F^S$ and $F^T$, whose 'mixed' relation with  $G^S$ and $G^T$
can be easily read off from eq.\VMIX\ itself (the equality holding
term by term).

In the present case, we modify the definition \ans\ of the $\theta$ angle
(and the phases) in the following way (consistently with eq.\VMIX):
\eqn\anss{
\eqalign{
 {1 \over Y}( F^S - { {\delta _{GS}} \over {8\pi^2(T+T^*)} } F^T)
& =\ {\sqrt 3}C\ m_{3/2} e^{i\alpha _S}sin\theta  \cr
{{\sqrt 3} \over  {T+T^*} } (1- { {\delta _{GS}} \over {24\pi^2 Y} } )^{1/2}
\ F^T\  & =\ {\sqrt 3} C\ m_{3/2} e^{i\alpha _T}cos\theta \cr
}
}
Notice that eqs.\VMIX\ and \anss\ reduce to eqs.\VVV\ and \ans\
in the limit $\delta _{GS}=0$, as they should.

After computing again the soft terms, one finds that the
resulting expressions can in practice
be obtained from the previous formulae \mor , \aor, \gor\
by making the replacements \yyy\ and
\eqn\soy{
cos\theta \ \longrightarrow \ (1-{ {\delta _{GS}}\over {24\pi^2 Y} })^{-1/2}
\ cos\theta
}
(without changing $sin\theta$).
The same happens with the formulae for the $B_{\mu}$ parameter.

We anticipate that the corrections due to $S-T$ mixing are normally
negligible for not too large  $\delta _{GS}$. However they turn out
to be important in the $\sin\theta \rightarrow 0$ limit in which all
tree level masses become small and one-loop effects cannot be neglected.

Notice that now  $F_S$ is non-vanishing when $\sin\theta \rightarrow  0$,
differently from the case without $S-T$ mixing (see eq.\ans).
However, if $\delta _{GS}$ is not too large, this limit still corresponds
to a moduli dominated SUSY-breaking.
Indeed, from eq.\anss\ one obtains
\eqn\TTSS{
|{{F^S} \over{F^T}}| = {|{\delta_{GS}}| \over{8\pi^2 (T+T^*)} } << 1  \ .
}
Let us finally comment that in specific orbifold models $\delta _{GS}$ is
an integer coefficient which is related to the cancellation of the
duality anomalies in the model (see refs.\DFKZ,\LO). In our present treatment
$\delta _{GS}$ will be taken more as a free parameter which {\it measures the
amount of one-loop mixing between the $S$ and $T$ fields} in the K\"ahler
potential. This type of mixing is expected to be present in generic
four-dimensional strings and not only in the orbifold case.

\subsec{Properties of soft terms independent of the cosmological constant}

The above analysis has shown that the different SUSY-breaking soft terms
have all an explicit dependence on $V_0$, i.e. the cosmological constant,
which is contained in $C$. In  section 6 we will discuss how to
face this fact. However, before that, it is interesting
to see if there are general properties which are {\it
independent of  the value of the cosmological constant} $V_0$.
For example, from the comparison of eqs.\AAF\ and \MMM\ one immediately
notices that the quantity $A/M_a$ is independent of the value of $V_0$,
since the factor $C$ drops in the ratio.
In particular, one gets
\eqn\aam{
{{A}\over {M_a}}\ =\ -1
}
in the $sin\theta =1$ limit  or (for any $\theta$) in the large $T$ limit
of Calabi-Yau compactifications, up to small threshold effects (see
eqs.\mag\ and \socy, respectively). In the  $sin\theta =1$ limit,
again ignoring small threshold effects and phases (see section 4),
one can also write a sum rule for scalar and
gaugino masses which is independent of $V_0$:
\eqn\sur{
M_a^2\ =\ 2 m_{3/2}^2\ +\ m_i^2
}
This formula tells us that, independently of the value of $V_0$, the gaugino
masses (at the string scale) are necessarily larger than the scalar masses.
Similar sum rules can also be written in the more general case with
$sin\theta \not= 1$, although some model dependence is necessarily
present.
 For the specific case of  the large $T$ limit of Calabi-Yau models
and for the orbifold-like models at tree level one finds
(see eqs.\socy\ and \mor,\gor\ respectively)
\eqn\mli{
M_a^2\ (1\ +\ {{3+n_i}\over 3}\ cotan^2\theta\ )\ =\ 2m_{3/2}^2\ +\ m_i^2\ .
}
up to small threshold effects and phases.
Notice how this formula collapses to eq.\sur\ when $sin\theta \rightarrow 1$.
The additional factor multiplying gaugino masses in this expression makes
possible the existence in this case of gaugino masses smaller than
squark masses.

\newsec{Complex phases in soft terms and the electric dipole moment of the
neutron}

The general soft terms obtained in the previous analysis
are in general complex.
Notice that if $S$ and $T$ fields acquire complex vacuum expectation
values, then the phases $\alpha _S,\alpha _T$ associated
with their auxiliary fields can be non-vanishing and
the functions $\omega _{ijk}(T,T^*), \Delta (T), Z(T,T^*)$ etc.
can be complex.

If one concentrates in the case of the MSSM, there are in general complex
phases
 in the
parameters $A$, $M$, $B$ and $\mu $, but two of them can be redefined away
(see e.g. ref.\DGH\ for a discussion of this point). In any phase convention
the phases
\eqn\fia{
\phi _A\ =\ arg(AM^*)\ \ ;\ \ \phi _B\ =\ arg(BM^*)\ \
}
cannot be removed. These phases, however, are quite constrained by limits on
the
electric dipole moment of the neutron (EDMN), since they give large one-loop
contributions
to this CP-violating quantity. One has in fact\DGH
\eqn\lim{
\phi _A\ ,\ \phi _B \ \leq \ 10^{-3}
}
for sparticle masses close to the weak scale.

It is interesting to study what are the constraints on the complex soft terms
computed in the previous sections coming from eq.\lim . To start with let us
compute the value of the phase $\phi _A$ at the string scale using the
previously
found formulae. Notice that $A$ and $M^*$ are proportional to
(see eqs.\AAF, \MMM)
\eqn\prop{
A\ \propto \ (e^{-i\alpha _S}\ +\ e^{-i\alpha _T}{{\omega }\over {tan\theta }})
\ ;\ M^*\ \propto \ (e^{i\alpha _S}\ +\ e^{i\alpha _T}{{\tau }\over {tan\theta
}
})
}
where $\omega$ was defined in eq.\omo and $\tau $ is given by
$\tau ={ {f_a^T (G_T^T)^{-1/2} } \over  {f_a^S (G_S^S)^{-1/2} } }$.
Defining
\eqn\def{
\eqalign{
\omega =&|\omega |e^{i\delta _{\omega }} \ ;\ \tau =|\tau |e^{i\delta
_{\tau }} \cr
\gamma _{\omega }  =&\alpha _S -\alpha _T +\delta _{\omega } \ ;\
\gamma _{\tau } = \alpha _S -\alpha _T +\delta _{\tau } \ \cr }
}
one gets after some algebra
\eqn\fha{
tan{\phi _A}\ =\ {  { |\omega ||\tau |sin(\delta _{\omega }-\delta _{\tau
})\ +\
tan\theta(|\omega |sin\gamma _{\omega }-|\tau |sin\gamma _{\tau })  }\over
{   tan^2\theta + |\omega ||\tau |cos(\delta _{\omega }-\delta _{\tau })
+tan\theta (|\omega |cos\gamma _{\omega }+|\tau |cos\gamma _{\tau })  } \
. } }
Now, eq.\lim\ tells us that $tan\phi _A\simeq \phi _A\leq 10^{-3}$. Barring
accidental
cancellations, it is clear that one can make $\phi _A$ arbitrarily small by
going to
a $tan\theta $ sufficiently large. In particular, for $tan\theta \sim 10^3$,
$\phi _A$ will be well within the experimental limits. This means that
{\it if SUSY-breaking comes essentially from the dilaton sector
( $sin\theta =1$) no
physical complex phases $\phi _A$ will be generated}, as it is obvious
from eqs.\AAF, \MMM.

Eq.\fha\  is substantially simplified in some limits. In particular, we know
that
a non-vanishing $|\tau |$ is only generated at one-loop
whereas $|\omega |$,
if non-vanishing, appears already at tree level. Then, as long as
$|\omega |\gg |\tau |$ (and $|\tau |\ll 1$ ) one has
\eqn\fhb{
tan\phi _A\ \simeq \ {  {|\omega |sin(\alpha _S-\alpha _T+\delta _{\omega })  }
\over  {tan\theta \ +\ |\omega |cos(\alpha _S-\alpha _T+\delta _{\omega }) } }
}
One sees that for large $tan\theta $ one has $\phi _A\sim |\omega |/tan
\theta $ and
one does not need to go to very large $tan\theta $ as long as $|\omega |$ is
small.
An example of this situation is the large $T$ limit of Calabi-Yau models
considered in the previous section. In that limit $|\omega |\rightarrow 0$ and
so does
$\phi _A$. For the orbifold models considered in the previous section
one does not expect a vanishing $|\omega |$ (up to particular
cases, e.g. fields in the untwisted sector, where  $|\omega |=0$,
see eq.\oor)
 and hence, barring accidental
cancellations, a very large $tan\theta $ is the simplest way to obey the
experimental limits.

However, the EDMN gets also equally important contributions from a
non-vanishing
 $\phi _B$
phase. Here, the situation is more complicated since, as we discussed in
section
 2,
there are many ambiguities in the determination of the $B$-parameter, depending
on
the mechanism assumed for the generation of a $\mu $-term.
It seems however that, in the case of $\phi _B$, obeying the experimental
bound is more complicated. In the
case of $\phi _A$, in the large $tan\theta $ limit what happens is that the
phases
of $M$ and $A$ coincide and $\phi _A\rightarrow 0$. However, in the case of
$\phi _B$,
the phases of $B$ and $M$ do not in general coincide, even  in the large
$tan\theta $ limit
(see e.g. eqs.\bch,\bzc\ and \mag). Thus $\phi _B$ will normally be large
unless
$\alpha _S$ and $\alpha _T$ are small (modulo $\pi $). We conclude that
{\it dilaton-dominated SUSY-breaking does not guarantee
a sufficient supression of the EDMN in the MSSM} unless further
assumptions about the origin of SUSY-breaking are made.

There is one interesting case in which dilaton-dominated SUSY-breaking
leads to a natural supression of the EDMN. Indeed, if the
$\mu $-problem is solved by the addition of a singlet $N$ as described in
subsection 2.2, the role of $B$ is played by $A_{\lambda }$ whose phase is
aligned with that of the gauginos and naturally
${\phi }_A={\phi}_{A_{\lambda}}=0$. Another possibility \ILCP\ is that
${\alpha _S}=0$ to start with. This may be a natural situation if
an $SL(2,Z)$ duality associated to the $S$ field were present in the
theory \FILQS.

In view of the above analysis it seems reasonable to impose in what
follows $\alpha _S=\alpha _T=0 \ mod\  \pi $.  Equivalently, we will set
to zero those phases and allow $\theta$ to vary in a range  $[0,2\pi )$.

\newsec{Some comments on soft terms and flavour-changing neutral currents.}

In section 2 we briefly mentioned the issue of flavour changing neutral
currents in string models. We now want to discuss several aspects of
this problem within the general approach here considered.

There are
several possible sources of FCNC in this scheme. To start with, if the
kinetic function of the matter fields is not diagonal (i.e.,
${\tilde K}^i_j\not= {\tilde K}_i \delta _{ij}$), after diagonalizing the
kinetic terms there will be e.g. non-diagonal $(mass)^2$-terms like
$\delta m^2_{{\tilde s}{\tilde d}}$. These type of terms give rise to
contributions to $K^0\leftrightarrow {\bar K}^0$ transitions,
$\mu \rightarrow e\gamma $ and other FCNC effects (see     \HAG \
and references therein for experimental constraints). There are three
comments in order        concerning this source of FCNC effects.
The first is that the experimental constraints are not terribly tight and
only affect in a significant manner the ${\tilde d}, {\tilde s}, {\tilde
e}$ and ${\tilde \mu }$ masses. Thus, e.g., from $K^0\leftrightarrow
{\bar   K}^0$ transitions one gets bounds    $\delta m^2_
         {{\tilde s}{\tilde d}} / m^2_{\tilde q}  \leq  10^{-2}$
typically (see ref.\HAG ). The second remark is that, of course, the
relevant mass terms are the low-energy ones, not the off-diagonal
terms generated at the string scale. One has to do the low-energy
running of the scalar masses and, for the squark case,
 for gluino masses heavier (or of the
same order) than the scalar masses, there are large flavour-independent
gluino loop contributions which are the dominant source of
scalar masses. We estimate this effect in a specific example below.
Thus the effect of primordial non-diagonal scalar masses at the
string scale is substantially diluted when taking into account the
running if gluino masses are not suppressed, as it happens in many string
scenarios (see below).
Finally, the third remark is that it is not difficult to find
string models in which ${\tilde K}^i_j$  is diagonal. That is the case,
for example, in generic orbifold compactifications. In these type of
models each chiral sparticle has different charges with respect to
enhanced $U(1)$ symmetries of the theory and off-diagonal kinetic
terms are forbidden. It is true that such off-diagonal terms can be
generated by non-renormalizable terms involving singlets in the theory.
However these will be suppressed by powers of the singlet VEVs over
$M_{Planck}$. Since, as remarked above, the experimental constraints
do not necessitate large suppressions we think that orbifold type models
provide examples with approximately diagonal kinetic terms.

Even if the mass terms induced at the Planck mass are flavour-diagonal,
there can appear FCNC effects from lack of universality of
soft masses for the different squark-slepton flavours. This is due to
explicit failure of the SUSY version of the GIM mechanism. Again one
finds \BC\ bounds of the type, e.g., $|m^2_{\tilde d}-m^2_{\tilde s}|
/ m^2_{\tilde q} \leq 10^{-2}$.
In some string models indeed the soft masses of different particles
may be different \IL . For example, the soft masses in orbifold
models depend on the modular weight of each particle (see eq.(6.4) in
next section). It is obvious that, in order to supress    FCNC effects it
is enough to assume that the modular weights of particles with the same
$SU(3)\times SU(2)\times U(1)$ quantum numbers are the same  \IL\
(e.g., $n_{\tilde d}=n_{\tilde s}$). Indeed this is the case of the
orbifold scenarios discussed in the following sections.  On the other
hand this is probably not strictly necessary if one takes into account
the low-energy running and $sin\theta $ is not too small.

It is instructive in this respect to consider the explicit case
of orbifold models (taking $C=1$) and see how the experimental
limits constrain the parameters of the model (i.e., $\theta$) in the
case in which, e.g., $n_{\tilde d}\not= n_{\tilde s}$. Let us neglect
here the effect of one-loop string corrections to both the K\"ahler
potential and gauge kinetic function. Then eq.\mor\ gives
for the soft scalar masses
$m^2_i=m^2_{3/2}(1+n_icos^2\theta )$. If one includes the effect of the
low-energy running one finds an approximate expression (see eq.(7.4))
\eqn\fcr{
m^2_i\ =\ m^2_{3/2}(1+n_icos^2\theta +24 sin^2\theta )
}
where the last term gives the effect of the low-energy running.
Then one finds
\eqn\frr{
{{|m^2_{\tilde d}-m^2_{\tilde s}|}\over {m^2_{\tilde d}+m^2_{\tilde s}}}
\ \simeq \ {{|n_{\tilde d}-n_{\tilde s}|cos^2\theta }\over
{50-(48-n_{\tilde d}-n_{\tilde s})cos^2\theta}} .
}
For example, for the previous ratio to be $\leq 10^{-2}$ it is enough to have
$cos^2\theta \leq 1/3$ for $|n_{\tilde d}-n_{\tilde s}|=1$. In this case the
constraint on $\theta $ is not very strong and one does not
need to go to  \IL ,\KL\  a
        complete  dilaton-dominated SUSY-breaking ($sin\theta =1$) to
get sufficient supression of FCNC.

The above discussion shows that the possible presence of too large
FCNC effects in string (and, more generally, in SUSY) models is not
as endemic as it seems at first sight. The experimental constraints
are not too tight and string models with the required properties exist.
We will thus concentrate in this paper in string models with
diagonal ${\tilde K}^i_j$s and flavour-independent masses for
sfermions of different generations (although we will allow for
intra-generational non-universality in some examples).

\newsec{ The role of the cosmological constant in soft terms.}

As mentioned in subsection 3.3, we have to face the fact that
the different SUSY-breaking soft terms have all an explicit
dependence on $V_0$, i.e. the cosmological constant,
and do something about it.
We cannot just simply ignore it, as it is often done, since
the way we deal with the cosmological constant problem has a bearing
on measurable quantities like scalar and gaugino masses.

Three possibilities concerning the cosmological constant problem and its
implications on soft terms come then to mind. They correspond to different
attitudes concerning this problem taken explicitely or implicitely in
the literature:

i) {\it Imposing $V_0=0$ as a constraint.}
 One may assume that there is {\it some yet undiscovered dynamics which
guarantees that $V_0=0$} at the minimum. In our context this implies
that the dynamics of the $S$ and $T$ superfields is such that their
auxiliary fields break supersymmetry with vanishing cosmological constant.
This could come about e.g. if some non-perturbative dynamics generates
an appropriate superpotential $W(S,T)$ with this $V_0=0$ property. Indeed,
such type of superpotentials can be constructed (see ref. \CFILQ), although the
physical origin of them is certainly obscure. Anyway, if one adopts this
philosophy one must set $V_0=0$ (or, equivalently, $C=1$) in all the
expressions for the soft terms. This was the assumption considered in ref.\KL.

In the numerical computations presented in section 7 we will
assume  this philosophy and we will also neglect phases
(for the reasons described in section 4). Under these conditions
the soft parameters for the large $T$ limit of Calabi-Yau
compactifications become (see eqs.\socy,\bcy,\apel):
\eqn\sos{
\eqalign{
m_i^2\ &=\ m_{3/2}^2sin^2\theta \cr
M_a\ &=\ {\sqrt 3}{{k_a ReS}\over {Ref_a}}m_{3/2}sin\theta \cr
A_{ijk}\ &=\ -{\sqrt 3}m_{3/2}sin\theta \cr}
}
\eqn\bms{
B_{\mu }\ =m_{3/2}(\ -1\ -\ {\sqrt 3}sin\theta \ -\ cos\theta)
\ =\ A\ -\ m_{3/2}(1\ +\  cos\theta )
}
\eqn\bzs{
B_Z\ =\ 2m_{3/2}(1\ +\ cos\theta )\
}
and for orbifold compactifications (see eqs.\mor,\aor,\gor,\bsi\ and
\gii) they are:
\eqn\sio{
\eqalign{
m_i^2\ &=\ m_{3/2}^2(1\ +\ n_icos^2\theta) \cr
A_{ijk}\ &=\ -{\sqrt 3}m_{3/2}sin\theta \ -\ m_{3/2}cos\theta (3+n_i+n_j+
n_k) \cr
 M_a\ &=\ {\sqrt 3}m_{3/2}{{k_a ReS}\over {Ref_a}}sin\theta \ +\ m_{3/2}
cos\theta ({{B_a'(T+T^*){\hat {G_2}}(T,T^*)}\over
{32\pi^3Ref_a}} ) \  \cr
B_{\mu }\ &=\ m_{3/2}( -1\ -\ {\sqrt 3}sin\theta \ -\
 cos\theta (3+n_H+n_{\bar H})) \  \cr
B_Z\ &=\ m_{3/2}( 2-(n_H+n_{\bar H}) \cos\theta \ -\
[2+(n_H+n_{\bar H})] \cos^2\theta) \ . \cr
}
}
We recall that, in the orbifold case, the one-loop modification mentioned
at the end of subsection 3.2 (see eqs.\yyy,\soy) should be added.

ii) {\it Keeping a $V_0\not= 0$ .}
 One might think that the experimentally constrained {\it cosmological term in
present cosmology is not directly connected to the Particle Physics vacuum
energy $V_0$}. This happens, for example, in the scenario first proposed by
Hawking in which spacetime, although highly curved at short distances
(of order of the inverse Planck mass), looks flat in average at long
distances. If this is the philosophy adopted, one should not worry about
a non-vanishing $V_0$. On the other hand {\it one cannot ignore the $V_0$
pieces
in  the formulae for the soft terms}. This is an important difference with
respect to the previous case.

iii) {\it Ignoring $V_0$.}
This is in fact the option chosen by most of the existing work on
SUSY-breaking through gaugino condensation. What happens is the following.
If one has an explicit model (like gaugino
condensation\FILQ ,\FMTV,\CFILQ,\CCMD,\A ) in which the
non-perturbative interactions involving the $S$ and $T$ fields are known, one
can explicitely compute the scalar potential and, as expected, one finds
a non-vanishing (negative) $V_0$. There is no known way to avoid that.
Then, if one still insists in setting $V_0=0$ (or, equivalently, in ignoring
$V_
0$),
one is {\it tacitly} assuming that there are some extra particles beyond $S$
and
 $T$
and the charged fields that force (somehow)  $V_0$ to vanish. These extra
fields
necessarily contribute to SUSY-breaking (in fact they are the dominant source
of
SUSY-breaking in gaugino condensation models!).

Our treatment of the soft terms is equally valid for the first two philosophies
above, since we keep in general a non-vanishing $V_0$.
In fact, as we saw in subsection 3.3,  one can reach
some general conclusions about soft terms which are independent of $V_0$.
On the other hand, the third philosophy is not included
straightforwardly in our scheme since it tacitly assumes that $S$ and $T$
are not the only relevant fields in the SUSY-breaking process. That is why it
is not surprising that adopting e.g. the first philosophy leads
to different qualitative conclusions compared to those found e.g. in
gaugino condensation. In section 8 we will compare the two approaches.

\newsec{Radiative $SU(2)_L\times U(1)$ breaking and SUSY-spectra expected in
string models}

The formulae for soft terms corresponding to  vanishing cosmological
constant and phases written in section 6 may lead to different phenomenological
situations depending on e.g. the possible values of the modular weights of
the particles,
the ansatz for the $B$ parameter, etc. In this section we are going to consider
three generic scenarios which will essentially cover the different
 possibilities
for soft terms within the present approach. The first  will be called the
{\it CY-scenario} and it corresponds to the large $T$ limit of Calabi-Yau
models
described in section 3. The second will be called the {\it O-I scenario}
(orbifold
models of type I). In this scenario the soft terms correspond to the orbifold
formulae obtained in section 3 with the additional constraint that the modular
weights $n_i$ and the value of $ReT$
are fixed in order to fit the appropriate joining of gauge
coupling constants through large string threshold corrections. The third will
be called the {\it O-II scenario} (orbifold models of type II). In this last
scenario all modular weights of particles are taken equal to $-1$ and $ReT$ is
of order
one.
This is, in some respects, very similar to the CY scenario. The reason why
we study this third scenario is because in it one can evaluate the
one-loop corrections to the K\"ahler potential. It is also, as we will see,
the only scenario in which gauginos may be lighter than scalars.
 We describe the motivation of these three scenarios below.

An interesting input
in order to further constrain the form of soft terms is imposing
phenomenological
constraints from other sectors of the theory. In particular, the apparent
success of the
joining of gauge coupling constants at high energies seems a reasonable one.
This is not
automatic in string models because it is well known that, with the particle
content of the
MSSM, the couplings join around $M_X\simeq 3\times 10^{16}$ GeV, whereas the
natural
unification scale in string models is $M_{string}\simeq 0.5 \times g_{string}
\times  10^{18}$ GeV\K, where $g_{string}=(ReS)^{-1/2}\simeq 0.7$. Thus
unification
takes place at energies around a factor $\simeq 12$ smaller than expected in
string theory.
This could be due to one of the following reasons:

a) There are large string $threshold$ $corrections$ which explain the mismatch
between $M_X$ and $M_{string}$. In some sense, what would happen is that the
gauge coupling constants will cross at $M_X$ and diverge towards different
values
at $M_{string}$. These different values appear due to large one-loop stringy
threshold corrections. This possibility has been analyzed in detail in
refs.\ILR,\IL \ for the case of orbifold models. The scenario O-I
discussed below is based in this assumption.

b)  There are $extra$ $chiral$ $fields$ in the spectrum apart from the minimal
set in
the MSSM. These extra fields need to gain large masses at some intermediate
stage
of symmetry breaking in between the string scale and the weak scale\AEKN.
In the C-Y and the O-II scenarios discussed below, this will be the tacit
assumption taken. We will assume that very close to the string scale
there are some very massive fields which explain this mismatch
between unification scale and string scale. In numerical computations
of the low-energy spectrum we will assume only the particle content of
the MSSM to perform the runnings. This is equivalent to the above
assumption that the extra superheavy fields sit very close to
the string scale. Let us remark that the numerical influence of the
mismatch between $M_X$ and $M_{string}$ on the running of the soft
terms is anyway negligible.

\subsec{Large $T$ limit of Calabi-Yau compactifications (CY-scenario)}

This scenario essentially amounts to assuming a K\"ahler potential as in
eq.\cyt\ , i.e., to all sparticles having effective modular weight -1.
The soft terms $m_i^2$, $A$ and
$M_a$ are given by eq.\sos\ whereas  the ans\"atze for the $B$ parameter are
provided
by eqs.\bms\ and \bzs . Notice that one has the boundary conditions at the
string scale
\eqn\scy{
{{Ref_a}\over {k_a ReS}} M_a\ =\ -A\ =\
{\sqrt 3}m_i\ =\ {\sqrt 3} m_{3/2}sin\theta \ .
}
where the index $i$ runs over all the chiral scalar fields in the model.
Notice that the relationship among the different soft terms is the same as the
one
found for the dilaton-dominated SUSY-breaking limit ($sin\theta =1$) discussed
in
subsection 2.3. The main difference now is that the overall mass scale is set
by
$m_{3/2}sin\theta $ and $not$ just $m_{3/2}$. Thus those boundary condititions,
in the
CY-scenario, survive {\it even for a moduli-dominated SUSY-breaking ($sin\theta
\rightarrow 0$)}.

The dependence on $\theta$ of the soft terms $M_a$, $A$
and $m_i$ in units of the gravitino mass
is shown in fig.1a, along with $B_Z$ and $B_{\mu}$. As explained
at the end of section 4, we let $\theta$ vary in the range
$0 \leq \theta \leq 2\pi$.
       The slight non-degeneracy of gaugino masses  is due to the
assumed existence of heavy threshold corrections close to the string
scale.
The different $Re f_a$ ($a=1,2,3$) were
computed by running up to $M_{string}$ the three gauge
coupling constants, obtained  from the following experimental inputs:
\eqn\exper{
M_Z=91.175\ GeV \ , \ \alpha_3(M_Z)=0.125 \ , \ \alpha_{em}(M_Z)=1/127.9
\ , \ sin^2 \theta_W (M_Z)=0.23
}
We have also assumed the standard grand unification values of the Kac-Moody
levels $k_3=k_2=(3/5) k_1=1$.
Notice that the limit $sin\theta \rightarrow 0$ is delicate since in that limit
all
$M_a,A,m_i$ vanish, for fixed  $m_{3/2}$.
Indeed, for very small $sin\theta$ the gravitino mass $m_{3/2}$
decouples from the SUSY-breaking soft terms and may become much larger than
them.
On the other hand, one can hardly ignore in that limit the one-loop corrections
to
the K\"ahler potential and hence the validity of the boundary conditions
in eq.\scy \ is thus
in doubt. Unfortunately, the one-loop corrections to the K\"ahler potential in
the large $T$
limit of CY compactifications are unknown.
These loop corrections have been estimated  in the orbifold case. We will see
in
 the O-II
orbifold scenario that for very small $sin\theta $ the role of the loop
corrections becomes
important. In the case of the CY-scenario here considered we will content
ourselves
with limiting the study to not too small $sin\theta $.

Let us now discuss the predictions for the low-energy ($\sim M_Z$)
sparticle spectra in this CY-scenario.
We will impose as an important constraint that the $SU(2)_L\times U(1)$
symmetry
be radiatively broken in the usual way. We will assume the minimal particle
content
of the supersymmetric standard model (i.e. the MSSM)
and that the gauge couplings unify with large threshold
corrections in the way described above. There are several particles whose mass
is
rather independent of the details of $SU(2)_L\times U(1)$ breaking and is
mostly given by the boundary conditions and the renormalization group running.
In particular, in the approximation that we will use
(neglecting all Yukawa couplings except the one of the top), that is
the case of the gluino $g$, all the squarks (except stops and left sbottom)
$Q_L=(U_L,D_L),U_R, D_R$ and all the sleptons $L_L=(V_L,E_L),E_R$.
For all these particles
one can write explicit expressions for the masses in terms of the gravitino
mass
 and
$sin\theta $. For the gluino mass at the weak scale one has
\eqn\gcy{ M_g \equiv
M_3(M_Z)\ =\ {{\alpha _3(M_Z)}\over {\alpha _3 (M_{string})}}\ M_3(M_{string})\
=\
{\sqrt 3}{{\alpha _3(M_Z)}\over {\alpha _0}}\ m_{3/2}sin\theta \ \equiv
q_3 m_{3/2}
sin\theta .
}
where $\alpha _0\equiv (4\pi ReS)^{-1}$. Numerically one finds for the
coefficient
$q_3=5.4$.
For the scalar masses one has
\eqn\scyy{
\eqalign{
m_i^2(M_Z)\ &=\ m_i^2(M_{string})\ +\ 2\sum _{a=1}^3 \left(
\ {{C^i_a}\over {4\pi }}M_a^2(M_{string})\alpha _a(M_{string})F_a(t)\
\right) \cr
&=\ m_{3/2}^2sin^2\theta \ \left( 1\ +\ 6 \sum _{a=1}^3 {{C^i_a}\over {4\pi}}
 {{k_a^2 \alpha^3 _a(M_{string})}\over {\alpha^2_0}}F_a(t) \ \right)
\ \equiv m_{3/2}^2sin^2\theta p_i^2 \cr}
}
where $C^i_a, a=1,2,3$ are the quadratic Casimirs corresponding to each scalar
labelled
by $i$
($C=(N^2-1)/2N$ for $SU(N)$, $C=Y^2$ for $U(1)_Y$). The functions $F_a(t)$ are
given by
\eqn\fau{
F_a(t)\ =\ {1\over {\beta _a}}\left( 1-{1\over {(1+\beta _a t)^2}} \right)\ ;\
\beta _a \equiv {{\alpha _a(M_{string})}\over {4\pi }}b_a \ ;\
t\equiv log(M_{string}^2/M_Z^2)
}
where $b_a$ are the $\beta $-function coefficients. To these contribution for
the soft masses
one has to add the D-term contributions which are normally negligible compared
to the terms in eq.\scyy . These may be found in eq.(31) of ref.\ILM .
One can compute the coefficients $p_i$
in eq.\scyy\ and gets
\eqn\ppp{
p_{Q_L}\simeq 5.1\ ;\
p_{U_R}\simeq p_{D_R}\simeq 5.0\ ;\
p_{L_L}\simeq 1.7 \ ; \
p_{E_R}\simeq 1.3 \ \ .
}
The gluino and the squark and slepton masses are thus in the universal ratio
(at the $M_Z$ scale)
\eqn\rat{
M_g\ :\ m_{Q_L} \ :\ m_{U_R}\ :\ m_{D_R}\ :\ m_{L_L}\ :\ m_{E_R}\ \simeq\
1\ :\ 0.94\ :\ 0.92\ :\ 0.92\ :\ 0.32 \ :\ 0.24 \ \ .
}
Although squarks and sleptons have the same soft mass \scy, at low-energy
the former are much heavier than the latter because of
the gluino contribution to the renormalization of their masses (see eq.\scyy).
The masses of these particles versus the gluino mass are plotted, for future
comparison
with the other two scenarios, in fig.2a.
Actually, the plot corresponds
to scalar masses without D-term contributions. The latter
depend on $tan\beta \equiv {{<{\bar H}>} \over {<H>}} $,
which in turn depend on the details of  $SU(2)_L\times U(1)$ breaking.
However, as mentioned above, the inclusion of D-terms leads
only to small shifts. In particular, for the extreme case of maximum D-term
contributions (large  $tan\beta$), the only modification is that the
$E_L$ and $E_R$ masses get slightly shifted upwards (a few GeV)
whereas the $V_L$ mass gets lowered (even below  $E_R$,
if the gluino is lighter than 350 GeV).

The rest of the supersymmetric mass spectra are  more dependent on the
$SU(2)_L\times U(1)$
breaking process and the value assumed for the $B$ parameter. Indeed, the
CY-scenario is
a four-parameter model: $m_0\equiv m_{3/2}sin\theta $, $B$, the top-quark mass
$m_t$  and the $\mu $ parameter\footnote*{For  simplicity in
the notation
we drop the hat in $\hat{\mu}$ as well as in  $\hat{h}_t$
from here on (see subsection 2.2).}.
 One can eliminate one of them (e.g. $\mu $,
 which is the one we know the least) in terms of the others by imposing
appropriate
symmetry breaking at the weak scale. The value of the top mass is quite
constrained by
the LEP and CDF data so that $m_t$ is not a source of big uncertainty.
The parameter $m_0$ essentially
fixes the overall mass scale whereas $B$ introduces some source of uncertainty.
In
general,  $B$ is a model dependent function of $m_{3/2}, sin\theta, T$, etc.,
as
 discussed
in sections 2 and 3. One could leave $B$ as a free parameter in the analysis,
but we find
more interesting to display the results in terms of some reasonable ansatz for
the
$sin\theta $ dependence of the $B$ parameter. In the present CY-scenario there
is an
interesting possibility to generate a $\mu $ term which does not seem to be
there in the
orbifold case. This is the mechanism discussed in section 2 which requires the
existence of an extra piece eq.\giu\ in the K\"ahler potential. As we mentioned
in
section 2 and 3,
under reasonable assumptions this mechanism leads to a $B$ parameter \bzs
\eqn\bra{
B_Z\ =\ 2 m_{3/2} (1\ +\ cos\theta )\ .
}
As $\theta $ varies, the obtained $B$ parameter varies in the range
$(0 - 4) m_{3/2}$.  (We will comment later what happens when
$\mu$ is generated via the alternative mechanism, giving rise to a different
$B$ parameter ($B_{\mu}$).)
We have performed a numerical analysis of the sparticle
spectrum consistent with radiative $SU(2)_L\times U(1)$ breaking and the
boundary
conditions in eqs.\scy\ and \bra . This is a three-parameter ($m_{3/2},\theta $
and
$m_t$) model once one imposes consistent symmetry breaking (to eliminate $\mu
$)
{}.
Thus {\it we have traded the four free soft parameters ($M,m,A,B$) of the MSSM
by
the two parameters $m_{3/2}$ and $\theta $}  in this CY-scenario. We are
not going to review here the well known technology of the renormalization-group
running of soft terms and radiative weak-symmetry breaking which may be found
e.g. in
ref.\ILO,\ILM  (see \R\ for a recent review and references). We now
describe the general
patterns of the
spectra.

In fig.3a we have plotted the masses of the different sparticles as a function
of
$\theta $. Instead of fixing $m_{3/2}$ to some given value, we have preferred
to fix the gluino mass $M_g=500 $ GeV  (equivalent to fix
$m_0=m_{3/2}sin\theta $, see eq.\scy) which is
of more direct physical relevance. The result for other values of the gluino
mass
may be obtained by appropriately scaling the figure (except for the particles
with masses lighter or of order $M_Z$).
We have also fixed in this figure the value of the top Yukawa coupling $h_t$
(at $M_{string}$) to $0.7$, which gives values of the top-quark mass in the
range $130\ GeV\leq m_t\leq 185\ GeV$, corresponding to the LEP range.
Let us also comment that taking $h_t=0.7$ corresponds to assuming an
approximate equality of this Yukawa coupling to the gauge coupling.
In specific string models, the non-vanishing Yukawa couplings are
often equal to the gauge coupling constants (up to constants of order
one). Thus taking as a first try $h_t=0.7$ seems a good option. We will
comment about the dependence on $h_t$ below.

Notice first of all that the regions of dilaton (modulus) dominance correspond
to
$\theta =\pi /2,3\pi /2$ ($\theta =0, \pi , 2\pi $). The regions of $\theta $
close to
$0,2\pi $ cannot be reached because the lightest stop-quark gets a negative
mass
 squared
at some point. This happens because in those regions the large value of $B_Z$
(see fig.1a)
together with the requirement that the scalar potential be bounded below force
$\mu$ to be large too, thus producing a large mixing in the stop mass matrix.
Besides, although it is not visible in the figure, a
narrow region around $\theta =\pi $ corresponds to a situation with $m_{3/2}
\rightarrow \infty $ and hence should be taken with some care (one-loop string
corrections are probably relevant).

Most of the spectrum comes qualitatively in three
types of particles with masses of the same order: the gluino-squark group
 ($g$ and $q$), the slepton and light-ino group ($l$ and
$C_1, N_1, N_2$) and the heavy-ino and Higgs group ( $ C_2,N_3,N_4$ and
$H,A,H^+
$).
$q$ denotes all squarks except stops, $l$ all sleptons,
$C$ charginos and $N$ neutralinos.
We have represented the gluino-squark group and the slepton group
as two (relatively compact) strips on the vertical axis.
The gluino-squark group is around 3-4 times heavier than the slepton/light-ino
group
whereas the heavy-ino/Higgs group is in between the other two groups for
$\pi /3 \leq \theta \leq 5\pi /3$. For values of $\theta $ beyond those limits,
the heavy-ino/Higgs group becomes even heavier than the gluino, as a
consequence
of the $\mu$ growth explained above.
The case of the stops $t_1, t_2$ is special. They are split in mass above and
below the gluino/squark group and  for values $\theta \geq 5\pi /3 $ or
$\theta \leq \pi /3$ the lightest of them ($t_2$) may get a negative
mass squared, which prevents the limits $\theta \rightarrow 0,2\pi $
from being reached, as explained above.
Coming back to charginos and neutralinos, notice that $N_1$ is approximately a
$B$-ino, with mass $\sim M_1(M_Z)$. Moreover, the pair $C_1,N_2$ essentially
corresponds
to $W$-inos (${\tilde W}^{\pm}$ and ${\tilde W}^0$), with mass $\sim M_2(M_Z)$.
Finally, the triplet $ C_2,N_3,N_4$ approximately corresponds to the charged
and neutral Higgsinos, with mass $\sim \mu$. Since the degeneracy in the
latter two groups holds to a good approximation (less than $6\%$),
we have plotted a representative particle for each of them (concretely,
the charginos $C_1$ and $C_2$). A similar choice has been done for the almost
degenerate heavy Higgs group $H,A,H^+$ ($A$ is plotted).

Fig.3a also shows the (tree-level) mass of the lightest Higgs $h$, whose
behaviour essentially reflects its $tan\beta$ dependence. In turn,
the correlation between $tan\beta$ and $\theta$ is mainly due to the
$\theta$ dependence of the $B$ parameter. Notice that the $h$ mass
varies from zero to its (tree-level) upper bound $M_Z$. In particular,
the upper bound is reached for $\theta \sim \pi $, since in that region
$B_Z$ is small and consequently $tan\beta$ is large.
Notice also that in the dilaton-dominated SUSY-breaking
($\theta =\pi /2 , 3\pi /2$) the mass of $h$ is of order $60$ GeV
\footnote*{Variations of $h_t$ and $M_g$ lead, in this case,
to an upper bound  $tan\beta \leq 2.5$ which implies
an upper bound on the $h$ mass $m_h \leq 65 $ GeV
and the top-quark mass $m_t \leq 185 $ GeV.}.
However, as one goes away from that limit and the $T$ contribution
increases one may obtain larger values $m_h$ of order $90$ GeV (for
$\theta \sim \pi $) or much smaller (as for $\theta \sim \pi /6$).
Thus, within the present scenario, a very heavy
$h$ Higgs would be an indication that there is a {\it substantial
component of moduli-dominated SUSY-breaking}.
This conclusion still holds after including one-loop corrections to
the $h$ mass. The tree-level values would be shifted upwards,
the amount of shift mainly depending on the top and stop masses.
For example, taking for those masses the values in fig.3a,
the upper (lower) values of $m_h$ would be shifted upwards
by roughly 30 (50) GeV.

If a different value of $h_t$ is used, the pattern of the
supersymmetric spectrum
will remain essentially unchanged. A change in $h_t$ will produce
(apart from an obvious change in $m_t$) only slight variations
in the part of the spectrum more sensitive to the
electroweak symmetry breaking, where $h_t$ plays a role.
For example, the choice $h_t(M_{string})=0.3$ (which implies
$125\ GeV\leq m_t\leq 145\ GeV$) gives lower values for the $\mu$
parameter, implying lower masses (a few tens of GeV)
for the heavy-ino and Higgs group and a very small reduction in the
$t_1 - t_2$ splitting. Moreover, the range of allowed values of
$\theta$ gets a bit reduced compared to the previous case,
and the 'cut' is now due to the potential getting unbounded at $M_{string}$.
Finally, the $h$ mass behaviour is similar to the previous case, apart
from a small shift of the curve towards the right.

The computation of the low-energy SUSY spectra in one of the
dilaton dominated limits ($\theta =\pi /2$) has also been independently
studied by the authors of ref.\BLM (see also \DIM ). Their numerical
results seem to agree with our results for this particular limit, after
correcting the factor 1/2 for the gaugino masses that we mentioned in
subsection 2.3.

We have repeated the analysis for the case where the alternative mechanism
for generating $\mu$ is at work, i.e. we have used
$B_{\mu}$ \bms\ instead of $B_Z$:
\eqn\bmss{
B_{\mu }\ =m_{3/2}(\ -1\ -\ {\sqrt 3}sin\theta \ -\ cos\theta)\ =\
A\ -\ m_{3/2}(1\ +\ cos\theta )
}
The resulting spectrum is shown in fig.3a'. The comparison with fig.3a
shows a rough similarity, apart from a global distorsion which renders
the figure asymmetric. In particular,
the curves corresponding to the heavy-ino/Higgs group (lightest Higgs)
reach their minimum (maximum) for $\theta \sim 7\pi/4$ instead than
for $\theta \sim \pi$. This is a consequence of the different $\theta$
dependence of $B_{\mu}$ and $B_Z$, which in turn influences
the above masses. This
different behaviour is manifest in fig.1a: notice that, differently from
$B_Z$, $B_{\mu}$ vanishes in a point lying in the fourth quadrant.
Apart from this difference, one could be surprised by the different
behaviour of the two spectra for $\theta \rightarrow \pi$, where
$B_{\mu}$ and $B_Z$ both vanish. However they vanish
for fixed $m_{3/2}$ (and at different speed), whereas we are now keeping
$M_g$ fixed, corresponding to a divergent $m_{3/2}\sim (sin\theta)^{-1}$.
Taking into account all this in eqs.\bra\ and \bmss,
$B_{\mu}$ turns out to be non-vanishing in that limit, differently
from $B_Z$. As a further manifestation of the different $B$ ans\"atze,
notice that for $\theta=\pi /2 \ (3\pi /2)$ (dilaton dominated SUSY-breaking)
the (tree-level) mass of $h$ is 60 (60) GeV for $B_Z$
and 45 (80) GeV for $B_{\mu}$. This shows how the spectrum in
the two dilaton-dominated limits depend on the choice of ansatz for
the $B$ parameter.

\subsec{Orbifold models with large string threshold corrections (O-I scenario)}

Let us turn now to study the second scenario, which corresponds to abelian
orbifold
four-dimensional strings in which appropriate joining of coupling constants is
obtained through large string threshold corrections.
 In orbifold models one expects an exponentially suppressed $\Delta (T)$ for
large $T$.
In many models one will have in fact $\Delta (T)=\eta ^2(T)$ and the
evolution of the gauge coupling constants will be given by
                                                \DKLD,\L,\DFKZ
\eqn\ggg{
\eqalign{
{1\over {g_a^2(Q)}}\ &=\ {{k_a}\over {g_{string}^2}}\ +\
{{b_a}\over {16\pi ^2}}log {{M_{string}^2}\over {Q^2}}\ \cr  &-\
{1\over {16\pi ^2}}(b_a'-k_a\delta _{GS})log((T+T^*)|\Delta (T)|^2) \cr }
}
The unification mass scale $M_X$ at which two gauge coupling constants
become equal, or more precisely $k_a g_a^2(M_X)=k_b g_b^2(M_X)$, is
related to $M_{string}$ by
\eqn\jcc{
{{M_{string}^2}\over {M_X^2}}\ =\ ((T+T^*)|\Delta (T)|^2)^{\gamma }\ \ ;\ \
\gamma = {{(b_a'k_b-b_b'k_a)}\over {(b_ak_b-b_bk_a)}}
}
where the $b'$s were defined in eq.\bpr .

As mentioned at the beginning of this section, we want the ratio
of eq.\jcc\ to be of order $(12)^2$. This can be achieved in an orbifold
model with $\gamma < 0$ and a moderately large $ReT$ value. A systematic
analysis of this possibility was done in refs.\ILR\   and  \IL  , and it was
found that
correct results for gauge coupling unification can be obtained for restricted
values of the modular weights of the massless fields. In fact, assuming flavour
independence and considering only the overall modulus $T$, one finds \ILR\
that the simplest possibility including lowest modular weights $|n_i|$
corresponds to
taking the following values for the SM fields:
\eqn\mwg{
\eqalign{
n_{Q_L}\ =\ n_{D_R}\ =&\ -1\ \ ;\ \ n_{U_R}\ =\ -2\ \ ;\ \
 n_{L_L}\ =\ n_{E_R}\ =\ -3 \ \ ; \cr
n_H&\ +\ n_{\bar H}\ =\ -5\ ,\ -4  \cr }
}
The above values together with a $ReT\simeq 16$ lead
to good agreement  for $sin^2\theta _W$ and $\alpha _3$.
The modular weights in eq.\mwg\  give us an explicit model in order to compute
the
soft terms in eq.\sio, a model which is consistent with
appropriate gauge coupling unification. It is also interesting
because it provides us with an explicit model with
non-universal scalar masses and shows the general features of
models with some of the modular weights different from -1.
For the masses of scalar particles with modular weights $-1,-2$ and $-3$,
eq.\sio\ gives respectively
\eqn\mnd{
m_{-1}^2= m_{3/2}^2sin^2\theta \ ;\
m_{-2}^2= m_{3/2}^2(1-2cos^2\theta ) \ ;\
m_{-3}^2= m_{3/2}^2(1-3cos^2\theta ) \ .
}
The $A$ term which is relevant to radiative symmetry breaking is the one
associated to
the top-quark Yukawa coupling $A_t$.
 For the modular weights of eq.\mwg,  one gets from eq.\sio
\eqn\ato{
A_t\ =\ -m_{3/2}\left( {\sqrt 3}sin\theta \ +\ n_{\bar H}cos\theta \right)
}
In order to compute the gaugino masses from
eq.\sio \ we have to input
$ReT\simeq 16$, as estimated in refs. \ILR,\IL\ and the values of
$B_a',\ a=3,2,1$. With the modular weights in eq.\mwg \ one finds from
eqs.\for\ and \bpr
\eqn\bpo{
B_3'\ =\ -6\ -k_3\delta _{GS}\ ;\
B_2'\ =\ -8 (or \ -7)\ -k_2\delta _{GS}\ ;\
B_1'\ =\ -18 (or\  -17)\ -k_1\delta _{GS}
}
where the numbers outside (inside) parenthesis correspond to the case
$n_H+n_{\bar H}=-5 \ (-4)$.
As we mentioned in section 3, $\delta _{GS}$ is a model dependent (but gauge
group
independent) constant (a negative integer with the present notation). Putting
all
factors together one finally finds the numerical expressions (for
$n_H+n_{\bar H}=-5$):
\eqn\moi{
\eqalign{
M_3 \ &=\ 1.0\ {\sqrt 3}m_{3/2}\left( sin\theta \ -\ (6+\delta _{GS})\
2.9 \times 10^{-2}\ cos\theta \right)    \cr
M_2 \ &= \ 1.06 \ {\sqrt 3} m_{3/2} \left( sin\theta \ - (8+\delta _{GS})\
2.9 \times 10^{-2} \ cos\theta \right)  \cr
M_1 \ & =\ 1.18 \ {\sqrt 3}m_{3/2} \left( sin\theta \ - ({54\over 5}+
\delta _{GS})\  2.9  \times 10^{-2} \ cos\theta \right)  \cr}\ .
}
(for $n_H+n_{\bar H}=-4$ replace  the corresponding $B_2'$, $B_1'$).
Concerning
the soft $B$ parameter, eq.\sio\  would give us
\eqn\boi{
\eqalign{
B_{\mu }\ &=\ m_{3/2}(-1\ -{\sqrt 3} sin\theta \ + \ cos\theta )
\ \ \ for \ n_H+n_{\bar H}=-4 \cr
B_{\mu }\ &=\ m_{3/2}( -1\ -{\sqrt 3} sin\theta \ + \ 2cos\theta)
 \ \ \ for \ n_H+n_{\bar H}=-5 \cr}
}
We recall that a $B_Z$-like parameter does not seem to appear in orbifold
compactifications.
If one wants to include the one-loop corrections to the K\"ahler potential
($S-T$-mixing) discussed in subsection 3.2 one just has to do the replacement
indicated in
eqs.\yyy\ and \soy. In fact, both the one-loop correction \soy\
and the one-loop gaugino term
proportional to $cos\theta $ in eq.\moi\ turn out to be numerically irrelevant
because,
due to  other consistency constraints, the allowed values of $sin \theta $ are
always
relatively close to one. The origin of this is eq.\mnd : for $cos^2\theta > 1/3
$ the
mass squared of the scalars of modular weight $-3$ (always present in this
scenario)
would become negative. Thus in order to avoid running into difficulties,
{\it consistent SUSY-breaking in the O-I scenario is confined to the regions
where
$sin^2\theta \geq 2/3$ }.
This means that in this case {\it the dilaton is necessarily the dominant
source of SUSY-breaking}. Notice, however, that the moduli contribution
to SUSY-breaking is $not$ in general negligible.
With respect to the one-loop correction \yyy, which
only appears in gaugino masses, it turns out to be also irrelevant
due to the fact that $\delta _{GS}$ is usually small.
In all orbifold models considered up to now $\delta _{GS}$
is of the same order of magnitude as the $b'$-coefficients appearing in the
model.

For the sake of definiteness, in the following we will focus on the case
$n_H=-2,n_{\bar H}=-3$. Other possible choices do not lead to significative
modifications in the phenomenological results.
The dependence of the above soft terms in units of the gravitino mass
as a function of $\theta $ is shown in fig.1b.
The results are essentially independent of the choice of
$\delta _{GS}$ as discussed above, so we have taken $\delta _{GS}=0$ for
simplicity. Several comments are in
order. First of all the plot only extends to the two regions where
$sin^2\theta \geq 2/3 $,  since in the complementary regions
the scalars with modular weight $-3$ (the sleptons and the Higgs field
${\bar H}$ in the present case)
would get negative squared masses. For $| sin\theta | =1$ one
recovers the dilaton
dominated SUSY-breaking results. However, as one gets away from that value
important deviations
occur. The scalars with large (negative) modular weights get soft masses
substantially smaller
than those with e.g. $n_i=-1$. In particular, the sleptons are much lighter
than
 the
squarks already at the string scale for $|sin\theta| \simeq 0.8$.  The value of
the $A_t$
soft parameter varies quite a lot ($0\leq |A_t|/m_{3/2}\leq 3$) in the allowed
range of $\theta
$. On the contrary, the gaugino masses show mostly the dependence on
$sin\theta
 $
in eq.\moi\ and the one-loop piece is numerically not very important. The
dependence of
the $B_{\mu }$ ansatz for the $B$-parameter is also shown.

Let us now turn to the low-energy particle spectrum in this O-I scenario. As in
the case of
the previous scenario, we analyse first  the sector
of the spectrum  which is rather insensitive to the radiative
$SU(2)_L\times U(1)$-breaking condition. Unlike the case of the
former CY-scenario in which all three terms $M_a,A$ and $m_i$ scale, to a first
approximation, like $m_{3/2}sin\theta $, in the present case the soft terms are
not
universal  in their dependence on $sin\theta $, as it is obvious from eqs.\mnd
,\ato \ and
\moi. Therefore in fig.2b we plot the masses of the gluino, the squarks
(except stops and left sbottom) and the sleptons
versus the gluino mass for a fixed
value of $\theta$, differently from fig.2a which is valid for any $\theta$.
In particular, we have chosen the value $\theta =
arcsin{\sqrt{2\over 3}}$, lying at the left boundary of one of the allowed
regions.
For such value of $\theta$
the contribution of the moduli field $T$ to SUSY-breaking is substantial.
The results are qualitatively similar to
 those of the  previous scenario, in spite of the different set of soft
scalar masses, because the low-energy scalar masses are mainly determined
by the gaugino contributions. The only exception is the $E_R$ mass,
which only feels the small $B$-ino contribution.
In particular, the mass ratios now turn out to be
\eqn\ratt{
M_g\ :\ m_{Q_L} \ :\ m_{U_R}\ :\ m_{D_R}\ :\ m_{L_L}\ :\ m_{E_R}\ \simeq \
1\ :\ 0.95\ :\ 0.92\ :\ 0.92\ :\ 0.25 \ :\ 0.11 \ \ .
}
where the similarity with eq.\rat\ is clear (the results would be
still more similar for the other possible values of $\theta$).
Analogously to fig.2a, the masses plotted in fig.2b do not contain
the D-term contributions, which depend on the process of symmetry breaking
and anyway are very small. In particular, concerning the possible shifts
of the slepton masses, similar remarks as in the case of fig.2a apply.

Let us consider now the sector of the spectrum which is more dependent on the
$SU(2)_L\times U(1)$ breaking mechanism and the $B$-parameter.
In fig.3b we show the results obtained using the ansatz of the
second equation of  \boi\ for the $B$ parameter.
As in fig.3a, we have fixed the top Yukawa coupling at the value
$h_t(M_{string})=0.7$ and the gluino mass at the value  $M_g=500 $ GeV.
Also in this case part of the spectrum comes qualitatively in three
types of particles with masses of the same order: the gluino-squark group
($g$ and $q$), the slepton and light-ino group
($V_L,E_L,E_R$ and $C_1, N_1, N_2$) and the heavy-ino and
Higgs group ( $ C_2,N_3,N_4$ and $H,A,H^+$).
We have also simplified the representation using criteria similar to the ones
already used in fig.3a, exploiting approximate degeneracies or
$\theta$ independence of some particle masses. Here we have chosen to show
explicitly the $E_R$ mass because it exhibits a characteristic
behaviour, reflecting the analogous behaviour of the corresponding
soft mass at $M_{string}$ (see $m_{-3}$ in fig.1c). A similar situation
holds for $E_L$ and $V_L$, which are slighty heavier than $E_R$.

For the reasons already explained,
the particle spectrum in fig.3b is confined to the two regions
around $\pi/2$ and $3\pi/2$ where $sin^2\theta \geq 2/3 $.
The differences between the two parts are essentially due to the different
values which $B$ takes in each of them.
Notice that the situation in both regions is similar to the corresponding
ones in fig.3a', since $B_{\mu}$ is similar for the two cases
when $sin\theta$ is large (see  eqs.\boi\ and \bmss).
In particular, in the region around $\pi/2$,
$B$ is never small, so that $tan\beta$ is not large and e.g. the
lightest Higgs is lighter than 45 GeV (at tree-level), and in
particular is lighter than $N_1$. In the region around
$3\pi/2$, $B$ is smaller (it can even vanish), so that $tan\beta$
is larger and the lightest Higgs mass can reach its (tree-level)
upper bound $M_Z$, and is always heavier than $N_1$.
The change in $B$ also implies a change in $\mu$, which gets smaller.
As a consequence, the  $t_1-t_2$ splitting is
somewhat reduced and the heavy-ino and Higgs group is lighter
(and becomes accidentally degenerate with the lightest stop
$t_2$).

We finally add that, if a different choice for $h_t(M_{string})$ is made,
remarks similar to the previous case (C.Y.) apply. The SUSY
spectrum is essentially  not affected, with possible slight variations
for the heavy-ino and Higgs group and the $t_1 - t_2$ splitting.
Also, for low values of $h_t$ the regions of allowed $\theta$ values
can get further restricted since the potential at $M_{string}$
can become unbounded.

Summing up, the results for the spectra in this type of orbifold
scenario are not very different from those of the previous case
for similar goldstino angle $\theta $. These {\it qualitative}
similarities are due to the fact that we are confined to regions with
$sin^2\theta \geq 2/3$ for consistency requirements. More generally,
{\it for a model with any of the modular weights $\not= -1$, one is
restricted to a region $sin^2\theta \ge 1/2$ and hence to
dilaton-dominance} in order to avoid unwanted negative
squared masses for some squark or slepton.

\subsec{Orbifold models with small string threshold corrections and
all $n_i=-1$ (O-II scenario)}

We mentioned when talking about the first (CY) scenario how, as $sin\theta $
decreases,
one can ignore less and less the one-loop corrections to the K\"ahler
potential.
 Furthermore,
the results get more and more dependent on the form of the string threshold
correction
function $\Delta (T)$, which is still quite uncertain in the context of
Calabi-Yau type
of compactifications. On the other hand, as we mentioned already in section 3,
both the
one-loop threshold effects and the one-loop corrections to the K\"ahler
potential
are much better known in the context of orbifold 4-D strings. Thus it make
sense
to study the orbifold analogous of the CY-scenario, which will explicitely
provide us
with one-loop corrected expressions for the soft terms. These corrections will
turn out
to be crucial in the $sin\theta \rightarrow 0 $ limit. Recall also that
this $sin\theta \rightarrow 0$ limit is only accessible to models with
all modular weights $n_i=-1$, as we mentioned in the previous subsection.

In this third scenario, which we will call for  definiteness the O-II scenario,
we will
assume that all chiral fields have modular weight $-1$ (much like effectively
happens in the
large $T$ Calabi-Yau limit). With such choice of modular weights we know that
the string
threshold corrections {\it cannot} account for the joining of gauge couplings
at
 a scale
$\sim 3\times 10^{16}$ GeV, as we said above. Thus we will be tacitly assuming
that there is
some other effect (e.g., existence of further chiral fields in the spectrum
below the
string scale) which appropriately mends the joining of the coupling constants.
If that is the case, there is no need to artificially chose a large value for
$ReT$ and hence we will take in numerical computations values for $ReT$ close
to
 the
duality self-dual point, which is what one would normally expect in a duality
in
variant
theory. A value suggested by several gaugino condensation analysis is
$ReT\sim 1.2$\ \FILQ,\FMTV.

Let us now describe the form of the soft terms in this O-II scenario, starting
with the scalar masses. They can be obtained from eq.\sio\
after introducing the one-loop correction from eq.\soy .
One gets:
\eqn\olo{
m_i^2\ =\ m_{3/2}^2\ (1\ -\ (1-{{\delta _{GS}}\over {24\pi ^2 Y}})^{-1}
cos^2\theta )
}
where $Y$ was defined in eq.\yyy . This result is numerically very similar to
that
in the CY-scenario as long as $sin\theta $ is not much smaller than one.
However, in the $sin\theta \rightarrow 0 $ limit in which the tree-level
scalar masses vanish one finds
\eqn\lil{
m_i^2\ (sin\theta \rightarrow 0)\ \simeq \
m_{3/2}^2\ ({{-\delta _{GS}}\over {24\pi^2 Y}})\ \simeq\ m_{3/2}^2 (-\delta
_{GS})
10^{-3}
}
where $\delta _{GS}$ was defined in section 3. In all orbifolds considered up
to
 now it is
a negative integer of the same order of magnitude as the $b'$-coefficients
appearing in
the model. We thus observe that, in the case of orbifolds, the inclusion of the
one-loop corrections in the K\"ahler potential has the effect of "regulating"
in
 some
way the $sin\theta \rightarrow 0$ limit yielding a non-vanishing result
for the
scalar masses. Concerning the $A_t$ coupling, since all modular weights are
equal to $-1$,
the associated fields will be untwisted (twisted associated to unrotated planes
of the underlying six-torus)
and the Yukawa coupling $h_t$ will be a constant (tend exponentially to
a constant).
Thus one has $\omega _t=0$ and a soft parameter (see eq.\sio):
\eqn\alo{
A_t\ =\ -{\sqrt 3} m_{3/2}sin\theta
}
just as in the tree-level result. Finally, concerning the gaugino masses,
they can be obtained again from eq.\sio\ after introducing the one-loop
correction from eqs.\yyy,\soy. One gets
\eqn\carlos{
 M_a\ =\ {{k_a Y}\over {2 Ref_a}} {\sqrt 3}m_{3/2}
(sin\theta \ +\ {{B_a'(T+T^*){\hat {G_2}}(T,T^*)}\over
{ 16{\sqrt 3}\pi^3 k_a Y}}  (1-{{\delta _{GS}}\over {24\pi ^2 Y}})^{-1/2}
cos\theta )
}
Since all modular weights are equal to $-1$, one has $b_a'=b_a$ (see eq.\bpr)
and one gets
\eqn\blo{
B_3'\ =\ -3\ -k_3\delta _{GS}\ ;\
B_2'\ =\ +1 \ -k_2\delta _{GS}\ ;\
B_1'\ =\ +11 \ -k_1\delta _{GS}
}
Then, using a value $ReT\simeq 1.2$, one obtains the following
numerical results
\eqn\mli{
\eqalign{
M_3 \ & \simeq \ 1.0\ {\sqrt 3}m_{3/2}\left( sin\theta \ -\ (3+{\delta _{GS}})\
4.6\times 10^{-4}\ cos\theta \right)    \cr
M_2 \ & \simeq \ 1.06 \ {\sqrt 3} m_{3/2} \left( sin\theta \ - (-1+{\delta
_{GS}
})\
4.6\times 10^{-4}\ cos\theta \right)  \cr
M_1 \ &  \simeq \ 1.18 \ {\sqrt 3}m_{3/2} \left( sin\theta \ - ({-33\over
5}+{\delta _{GS}})
\  4.6\times 10^{-4} \ cos\theta \right)  \cr}
}
Notice that now the one-loop correction to gaugino masses is much
smaller than in the
previous scenario \moi\ because the value of $ReT\simeq 1.2$ is much smaller
than in the
previous case ($ReT\simeq 16$). The one-loop piece has essentially a linear
dependence
on $ReT$ for large $T$. Concerning the ansatz for the $B$ parameter, eq.\sio\
after introducing the one-loop correction according to eq.\soy\
yields
\eqn\blo{
B_{\mu }\ =\  m_{3/2}(-1\ -{\sqrt 3} sin\theta \ -\
(1-{{\delta _{GS}}\over {24\pi^2 Y}})^{-1/2} cos\theta) \ \ .
}
It is obvious from the above equations that for not too small $sin\theta $ the
results are
almost identical to those found in the CY-scenario \scy.
Thus we will concentrate in
analysing in detail what happens in this scenario for the $sin\theta
\rightarrow
 0$ limit,
which we do not know how to explore in the CY case. We recall that this limit
corresponds to a moduli dominated SUSY-breaking even if $S-T$ mixing is
present, as discussed at the end of subsection 3.2. We are really interested in
understanding the qualitative behaviour of that limit and hence we will just
take a fixed
value for $\delta _{GS}$, e.g. $\delta _{GS}=-5$. This is a negative integer
with a
magnitude of order of the $b'$ coefficients involved and hence it is not an
unreasonable
value. We will comment below what happens as we vary this parameter.
Finally, although $sin\theta \sim 0$ both for $\theta \sim 0$
and $\theta \sim \pi$, we will focus on this latter case for reasons to be
explained below.

We show in fig.1c the soft terms in units of the gravitino mass as a function
of $\theta-\pi$, for $|\theta-\pi|\leq 0.1$, $ReT\simeq 1.2$ and
$\delta _{GS}=-5$. The most prominent feature of the soft
terms is that {\it for values of $|sin\theta |$ below $5\times 10^{-2}$ the
gaugino masses
become smaller than the scalar masses}. This is something which is
qualitatively
different from the results in e.g. the O-I scenario, where gaugino masses are
always necessarily larger than scalar masses.  On the other side,
this situation is quite similar to the one obtained in explicit
gaugino condensation models\CCM\  although
it is not really identical, as we will comment in the next section.
Indeed, in this O-II scenario
as $sin\theta $ decreases the gaugino/squark mass ratio decreases.

Notice that the qualitative behaviour found here for small $sin\theta $ is
generic
for any non-vanishing negative integer $\delta _{GS}$. The only difference
is the particular value of $sin\theta $ at which the gaugino masses start being
smaller than the scalar masses. Also, different values for $\delta _{GS}$ lead
to
different gaugino mass ratios (e.g. $M_3/M_2$) as $sin\theta \rightarrow 0$,
but
we consider this as a small correction to the most relevant feature found in
this
limit, which is that the gaugino masses become small compared to the scalar
masses. (The case
$\delta _{GS}=0$ is  special since, as can be seen from eq.\lil , the scalar
masses tend to
zero  and the gaugino masses provide essentially the only source of
SUSY-breaking for
$sin\theta =0$. However, generically there will be $S-T$ mixing
in the K\"ahler potential and a case with $\delta _{GS}=0$ is atypical.)
 Another point to remark is that in the present limit
the gravitino mass is much larger (more than an order of magnitude bigger) than
the soft masses. For example, for $\delta _{GS}=-5$, eq.\lil\ implies that
$m_{3/2} \simeq 14 m_i$.

Let us describe now what is the structure of the low-energy SUSY spectra in
this
small $sin\theta $ limit of the O-II scenario. Let us start as usual with
the sector of the spectrum which is rather insensitive to the radiative
electroweak breaking, i.e. the gluino, the
squarks (except stops and left sbottom) and the sleptons.
Since the soft terms
have a different dependence on $\theta $, we will content ourselves with
showing results for a fixed small value of $sin\theta $, because this is the
limit we want to explore (for large $sin\theta $ the results correspond to
a good approximation with those of the CY-scenario).
Fig.2c shows the squark, slepton and gluino masses
(versus the gluino mass) for the illustrative choice
$\theta-\pi=5\times10^{-3}$
,
with $ReT\simeq 1.2$ and $\delta _{GS}=-5$.
The situation now is completely reversed with respect to fig.2a and fig.2b
which
correspond to the CY and O-I scenarios. The gluino is substantially lighter
than the scalars.  For example,
in this particular case the relation is $m_{q,l} \simeq
2.5 M_g$. Notice also that the physical masses of squarks and sleptons
are almost degenerate. This happens because the universality
of soft scalar masses at high energy is not destroyed by the
gluino contribution to the mass renormalization, which is now very small.

The above results tell us that {\it if gluinos lighter than squarks and
sleptons
 are found,
this could be an indication that the dominant source of SUSY-breaking lies in
the
moduli} and not in the dilaton sector.

Concerning the rest of the spectrum, we will use the expression \alo\ for
the $A_t$ parameter and the ansatz \blo\ for the $B$ parameter.
As in fig.3a and fig.3b, we will fix $h_t(M_{string})=0.7$ and
$M_g=500 $ GeV. Fig.3c shows the spectrum in a range
$0.003<\theta-\pi<0.03$, again for $ReT\simeq 1.2$ and $\delta _{GS}=-5$.
We focus on the case $\theta \sim \pi$ because in the other
region $\theta \sim 0$ the lightest stop has negative
mass squared due to the large value of $\mu$ \footnote*{We should add that, for
$\theta \sim 0$, the radiative breaking conditions give also
a second solution for $\mu$, provided one uses a smaller value
of $h_t$. The value of such $\mu$ is only a few GeV,
implying a particle spectrum with e.g. a too light chargino.}.
The most apparent feature of fig.3c
is the increase in mass of most particles when lower
values of $\theta-\pi$ (i.e. $sin\theta$) are approached.
The reason is that, in the region considered, such masses
are essentially proportional to $m_{3/2}$ (see e.g. \lil), which
is increasing because we are keeping $M_3$ fixed.
Notice also that the lightest
Higgs mass is always close to its (tree-level) upper bound $M_Z$,
due to the large $tan\beta$ induced by the small value of $B$.
A different initial value of $h_t$ leads to minor changes only,
similar to the ones commented in the previous cases (C.Y. and O-I).
Besides, if $h_t(M_{string}) \leq 0.3$ the region
$|\theta-\pi|\leq 0.01$ cannot be reached because the symmetry
breaking conditions have no solution.

\subsec{Discussion of the overall supersymmetric spectra}

Let us try to summarize the most prominent patterns obtained for the spectra
of supersymmetric particles in this large class of models. For a given
choice of string model the free soft
parameters of the MSSM ($M,m,A,B$) are {\it given in terms of the
gravitino mass $m_{3/2}$ and the goldstino angle $\theta $}. In some
string models in which the one-loop corrections become important
(O-II scenario) additional dependence on other parameters ($\delta _{GS},
ReT$) may appear, although the latter are less crucial in understanding the
{\it qualitative} patterns of soft terms.

One first point to remark is that one can have flavour-independent
soft scalar masses even without dilaton-dominated SUSY-breaking. In fact, all
the scenarios discussed above have flavour-independent scalar masses, although
in the case of the O-I scenario different scalars within the same
flavour generation can have different masses. Thus {\it dilaton-dominated
SUSY-breaking is a sufficient but not necessary condition to obtain
scalar mass universality}.

For goldstino angle $|sin\theta|\geq 5\times 10^{-2}$ the results for the
different scenarios are not terribly different. The heaviest particles are
the coloured ones with gluinos and squarks almost degenerate and $3$ to $6$
times heavier than sleptons and the lightest chargino ($C_1$) and the
next to lightest neutralino $N_2$. The heaviest neutralinos ($N_3,N_4$) and
chargino ($C_2$) along with the heavier Higgsses ($H,A,H^+$) oscillate in
between the masses of squarks and sleptons, depending on the precise value
of $sin\theta$. This variation is largely due to the $\theta $-dependence
of the $B$-term. In all cases the lightest supersymmetric particle is
essentially the $B$-ino. The lightest Higgs ($h$) tree-level mass varies in
all the allowed range from $0$ to $90$ GeV, depending on the value of
$sin\theta$.

For very small $sin\theta$ one can no longer neglect in general the one-loop
corrections. Now the situation concerning the spectrum is very much changed
and the gluino may be even lighter than the squarks and sleptons. The latter
become almost degenerate and the values of $tan\beta $ may grow quite large
leading to $h$-Higgs masses close to the upper bound.
If a spectrum of this type
is found, it could be an indication of a modulus-dominated SUSY-breaking.

It is also important to recall that values $sin^2\theta \le 1/2$ are
only possible in models in which all modular weights are equal to -1.
Otherwise some of the squarks and/or sleptons would get negative
squared masses at the string scale.

\newsec{A comparison with the gaugino condensation approach.}

Gaugino condensation effects in the hidden sector are able to break
SUSY \gcon \
     at a hierarchically small scale, at the same time as the dilaton
$S$ and the overall modulus $T$ acquire reasonable
 VEVs \FILQ,\FMTV,\CFILQ,\CCMD,\A. In particular,
$Re T$ is close to the duality self-dual point, $Re T \simeq 1.2$
\FILQ,\FMTV, and using
several gaugino condensates it is possible to obtain
$g_{string} =(Re S)^{-1/2} \simeq 0.7$\CCMD. This
was realized in the context of orbifold constructions, where also the
soft SUSY-breaking terms were studied\CFILQ,\IL,\CCM,\KL. In a
recent analysis \CCM\ the
soft terms have been calculated taking into account the one-loop corrected
K\"ahler potential (i.e. $\delta_{GS} \neq 0$). It turns out that
scalar masses tend
to be larger than gaugino masses. Since this generic conclusion was
also obtained in the O-II scenario of section 7, it is interesting to compare
our approach to that in ref.\CCM.

In the calculation of section 7 we assumed that some non-perturbative
dynamics generates an appropriate superpotential $W(S,T)$  giving vanishing
cosmological constant $V_0=0$. This is not possible in the gaugino condensation
mechanism since the non-perturbative interactions involving the $S$ and $T$
fields are known and they originate a non-vanishing (negative) $V_0$. If
one insists in setting $V_0=0$ (or, equivalently, in ignoring $V_0$), one
is tacitly assuming some extra particles beyond $S$, $T$ and the charged
fields that force (somehow) $V_0$ to vanish. Therefore, one should add
in the VEV of the scalar potential \VMIX\ a contribution
due to the auxiliary fields associated to the new particles.
For simplicity we summarize this contribution in the form $G_A^A |F^A|^2$,
corresponding to the case of a unique (extra) field $A$ which does
not mix with any other field of the theory.

The fact that now there are three fields ($S,T,A$) which play a role in the
process of SUSY-breaking forces us to define two {\it goldstino angles}
(consistently with $V_0=0$):
\eqn\ansss{
\eqalign{
 {1 \over Y}( F^S - { {\delta _{GS}} \over {8\pi^2(T+T^*)} } F^T)
& =\ {\sqrt 3}\ m_{3/2} e^{i\alpha _S}sin\theta cos\theta_A  \cr
{{\sqrt 3} \over  {T+T^*} } (1- { {\delta _{GS}} \over {24\pi^2 Y} } )^{1/2}
\ F^T\  & =\ {\sqrt 3} \ m_{3/2} e^{i\alpha _T}cos\theta  cos\theta_A \cr
(G_A^A)^{1/2} F^A\  & =\  {\sqrt 3}\ m_{3/2} e^{i\alpha _A}sin\theta_A \cr
}
}
Now it is straightforward to obtain the modified soft scalar masses
\eqn\ollo{
m_i^2\ =\ m_{3/2}^2\ (1\ +\ n_i (1-{{\delta _{GS}}\over {24\pi ^2 Y}})^{-1}
 cos^2\theta \ cos^2\theta_A )
}
Of course in the limit $sin\theta_A=0$  we recover eq.\anss\ and therefore
 the expression \olo\ of the O-II scenario (with $n_i=-1$).
However, in the gaugino condensation mechanism the form
of $W(S,T)$ is completely determined and therefore at the minimum of the
scalar potential the value of the angles can be calculated. In particular
one finds $cos\theta \simeq 1$ and $cos\theta_A \simeq 0.28$.
Using these results in eq.\ollo\  we obtain
\eqn\olllo{
m_i^2\ =\ m_{3/2}^2\ (1\ +\ n_i (1-{{\delta _{GS}}\over {24\pi ^2 Y}})^{-1}
0.078 ) \simeq  m_{3/2}^2
}
This equation\CCM\ can be compared with \lil\ of the O-II scenario where
$m_i<<m_{3/2}$. The qualitative difference is clear: in the gaugino
condensation approach $m_{3/2} \leq O(1 TeV)$ in order to fulfil the
naturalness
bounds on scalar masses whereas in the O-II scenario $m_{3/2}$ is much larger
(for instance, the explicit example analysed in subsection 7.3 gives
$m_{3/2} \simeq 14 m_i$).

The gluino mass can also be explicitely calculated in the gaugino
condensation approach and its relation
with scalar masses at low-energy is \CCM
\eqn\gcond{
{{m_i} \over  {M_g}} \simeq {1000 \over {(-\delta_{GS})}}
}
{}From this equation it is straightforward to obtain a lower bound
on $\delta_{GS}$, taking into account naturalness bounds on scalar
masses and experimental limits on gluino mass ($M_g > 135$ GeV):
$(-\delta_{GS})\geq 30$. On the other hand, an upper bound on the ratio
${{m_i} \over  {M_g}}$ can also be obtained since  $(-\delta_{GS})$
is in general a not very large number, as discussed in the previous
section. For example,  the extremely conservative bound
$(-\delta_{GS}) \leq 100$ gives ${{m_i} \over  {M_g}} \geq 10 $.

These are other qualitative differences with the O-II scenario, where
$(-\delta_{GS})$ and ${{m_i} \over  {M_g}}$ can be smaller than 30 and
10, respectively (for instance, in the example discussed in subsection
7.3, $\delta_{GS}=-5$ and  ${{m_i} \over  {M_g}} \simeq 2.5$).
If in the future gluinos and scalars are detected and they
fulfil $1 \leq {{m_i} \over  {M_g}} \leq 10 $, then the gaugino
condensation scenario would be discarded in favour of the O-II one.

Notice that both in our previous formalism and in this extension to
cover for models with extra chiral fields contributing to
SUSY-breaking, the primordial gaugino mass verifies $M\le {\sqrt 3} m$
whenever all modular weights are $-1$.
The addition of extra contributions always goes in the direction of
decreasing gaugino versus scalar masses. The reason is obvious: at tree
level only $F_S$ contributes to gaugino masses whereas scalar masses are
in general sensible to all the non-vanishing auxiliary fields present. Loop
corrections only slightly modify  this situation.

\newsec{Final comments and conclusions.}

In the present paper we have tried to provide for a string-motivated
parametrization of the SUSY-breaking mass parameters of the
supersymmetric standard model. The idea is similar in spirit to
what we do in the standard model in which we parametrize our
ignorance of the process of $SU(2)_L\times U(1)$ breaking by
assuming that an
 (elementary or composite) Higgs field  acquires
a non-vanishing vacuum expectation value. Here we assume that some
auxiliary fields of some chiral multiplets acquire VEVs breaking
supersymmetry spontaneously, without specifying what is the origin of
SUSY-breaking. The key point is to identify
what fields are involved in the process.

Four-dimensional strings
provide for natural candidates to play an important role in SUSY-breaking:
the dilaton and moduli fields which are perturbatively massless and are
generically present in large classes of string models. In the simplest
scenario SUSY-breaking is assumed to take place through the chiral
superfields $S$ and $T$ associated to the complex dilaton and the
overall modulus of the compact space respectively. We find useful to
introduce a "goldstino" angle $\theta $ such that
$tan\theta =|F_S|/|F_T|$, thus telling us where the dominant
source of SUSY-breaking resides. For $sin\theta =1$ one has dilaton dominance
whereas for $sin\theta \rightarrow 0$ one has modulus dominance. All
formulae for soft terms get substantially simplified when written in
terms of this goldstino mixing angle. For a $given$ string  model
the free soft parameters of the MSSM ($M, m, A, B$) are essentially traded
for two parameters $m_{3/2}$ and $sin\theta $ (see e.g. formulae
(6.1)-(6.4)).

After providing some formulae of more general validity,
we have computed the
soft terms for some classes of string models, particularly for the large
radius limit of Calabi-Yau type of models and for orbifold models.  We
have kept all the way the possible complex phases which may arise in the
 process of SUSY-breaking. These are relevant in providing complex
phases for the soft terms which may be important in CP-violating
phenomena. The experimental limits on the electric dipole
moment of the neutron (EDMN) put severe constraints on those phases.
We show that, assuming the particle content of the MSSM,
dilaton dominance ($sin\theta =1$) is not enough to guarantee
enough supression of the EDMN. We briefly discussed some other possible
alternatives. We also briefly commented upon the constraints coming
from experimental limits on flavour changing neutral currents (FCNC).
We argued that those constraints are interesting but not difficult
to avoid. One of the results one obtains (already remarked in \IL ) is that one
can get universal soft masses for all the scalars without
assuming dilaton-dominated SUSY-breaking.

We have also pointed out that the treatment one gives to the
cosmological constant problem has an impact on the resulting
soft masses. If one assumes that the SUSY-breaking dynamics
cancels the cosmological constant, the results obtained for the
 soft masses are {\it different} than those obtained computing
the soft terms at the minimum of a scalar potential with
non-vanishing vacuum energy. This explains why the results
obtained in the present analysis are different from those
obtained previously in explicit gaugino condensation models.
We argue that, in practice, the gaugino condensation models assume
the existence of extra fields beyond $S$ and $T$ with a
leading role in SUSY-breaking. We discussed in some detail the
relationship between our general approach and the gaugino
condensation models by introducing a second goldstino angle
which extends our formulation to include the possibility of
extra SUSY-breaking chiral fields beyond $S$ and $T$. We also
showed some mass sum-rules which are independent of the value
of the cosmological constant (subsection 3.3).

The paper contains a renormalization group analysis of the low-energy
spectra obtained from soft terms as computed in the text. This
includes also constraints coming from appropriate radiative
$SU(2)_L\times U(1)$ breaking. To do that we considered three type of
generic string scenarios called CY, O-I and O-II respectively.
The first (CY) corresponds to assuming identical "modular weights"
$n_i=-1$ for all squark, slepton and Higgs superfields. This is the
situation one has, e.g., in the large radius limit of Calabi-Yau type
compactifications. The third (O-II) assumes essentially the same but
includes one-loop string corrections to the K\"ahler potential and
gauge kinetic function as computed in orbifold models. The second
scenario (O-I) assumes different (flavour-independent) modular weights
 for the different squark and sleptons within each generation. The
modular weights are chosen so that one can have appropriate large
string threshold corrections to fit the joining of gauge coupling
constants at a scale $\simeq 10^{16}$ GeV (see refs.\ILR ,\IL ).

It turns out that models with some of the modular weights different
from -1 (like the O-I scenario) are forced to have
 $\sin^2\theta \geq 1/2$ in order to avoid getting negative
squared-masses for some of the squarks and/or
sleptons. Due to this fact one finds similar $qualitative$ results for
the low-energy spectrum of sparticles in all the above
string scenarios as long as $|sin\theta |\geq 5\times 10^{-2}$.
The heaviest particles are the coloured ones with squark and gluinos
almost degenerate and 3 to 6 times heavier than sleptons and the
lightest chargino and the next-to-lightest neutralino. The heaviest
two neutralinos and chargino along with all but the lightest Higgs
have masses oscillating in between the masses of squarks and sleptons,
depending on the precise value of $sin\theta $. This variation is largely
due to the  (model-dependent) $\theta $-dependence of $B$. In all cases
the lightest supersymmetric particle is almost purely the Bino.
The lightest Higgs (h) tree-level mass varies in all the allowed
range from 0 to 90 GeV, depending on the value of $sin\theta $.

For very small $\theta $ (modulus-dominance) the one-loop string
corrections turn out to be important. This limit is only accessible
if all modular weights of sparticles are equal to -1, as in the
CY and O-II string scenarios. Now the form of the spectrum
is completely changed and the gluino may be lighter than the squarks
and sleptons which are almost degenerate. This type of spectra is
qualitatively similar to those found in gaugino condensation models.
However, this similarity is accidental as remarked in section 8.
Furthermore, the ratio of gaugino versus scalar masses found
in gaugino condensation models turns out to be generally
much smaller than the
corresponding ratio in the present modulus-dominated scenario.

The general pattern of SUSY-spectra found in the present approach
are very characteristic.
 Optimistically, if the spectrum of
SUSY particles is eventually found, one will be able to rule out
(or rule in) some of the general scenarios (e.g., dilaton or
modulus dominance) here discussed. More modestly, we hope that
the formulae and the examples worked out in this paper will be
of help in looking for a more fundamental understanding of the
origin of SUSY-breaking soft terms in the supersymmetric standard
model.

\bigskip
\bigskip
\bigskip
{\bf Acknowledgements}

We thank A. Casas, J. Louis, D. L\"ust, F. Quevedo and G. Ross
for useful discussions. A. Brignole acknowledges the spanish Ministerio
de Educaci\'on y Ciencia for a postdoctoral fellowship.

\bigskip
\bigskip
\bigskip
\bigskip
\bigskip
\bigskip
\bigskip
\bigskip
\bigskip
\bigskip
\bigskip
\bigskip
\bigskip
\bigskip
\bigskip
\bigskip
\bigskip
\bigskip
\bigskip
\bigskip
\bigskip
\bigskip
\bigskip
\bigskip
\bigskip
\bigskip
\bigskip
\bigskip
\bigskip
\bigskip
\bigskip
\bigskip
\bigskip
\bigskip
\bigskip
\bigskip
\bigskip
\bigskip
\bigskip
\bigskip
\bigskip
\bigskip
\bigskip
\bigskip

{\centerline{\bf ERRATUM}}

\bigskip
{\centerline{\it Towards a theory of soft terms for the supersymmetric
standard model}}
{\centerline{\it Nucl. Phys. B422 (1994) 125}}

\bigskip
\bigskip

Among the formulae for the soft terms which appear in the paper,
the ones corresponding to the parameter $B_Z$ are not correct.
We recall that $B_Z$ is the $B$--parameter corresponding to one
of the possible mechanisms for generating the $\mu$--parameter.
Such mechanism is described in Section (2.2), point (d-iii).
However, the expression for the effective ${\hat \mu}$ given
in the text below eq.(2.17) is not complete, because a factor
$$
X \equiv 1- C{\sqrt{3}} e^{i \alpha_T} \cos\theta
({K_0}^T_T)^{-1/2} {{Z_T} \over {Z}}
$$
\noindent is missing. The correct ${\hat \mu}$ can be written synthetically as
${\hat \mu}=m_{3/2} Z X ({\tilde K}_H {\tilde K}_{\bar H})^{-1/2}$.
As a consequence the expression for $B_Z$ in eq.(2.20) should
be divided by $X$. In addition, the first term on the r.h.s.
should read $3(C^2-1)$, i.e. the denominator $K$ should be removed.

Some particularizations of eq.(2.20) appear in the following
sections. Therefore the same modifications should be made.
More explicitly:

Eq.(2.24) should read
$$
B_Z = m_{3/2} (2 + 3 (C^2-1))   \,\,  ;
$$

Eq.(3.6) should read
$$
B_z = m_{3/2} {{ 3(C^2-1)+2+ 2 C \cos\alpha_T \cos\theta} \over
{1+Ce^{i\alpha_T}\cos\theta}}  \,\,  ;
$$

The last formula in eq.(6.4) should read
$$
B_Z = m_{3/2} {{ 2-(n_H+n_{\bar H}) \cos\theta
-[2+(n_H+n_{\bar H})] \cos^2\theta} \over {1+\cos\theta}}  \,\, ;
$$

Eqs.(6.3) and (7.8) should read
$$
B_Z = 2 m_{3/2}   \,\,  .
$$
\noindent The last formula is by itself interesting because
$B_Z$ becomes independent of $\theta$, for fixed $m_{3/2}$.
The graph of $B_Z$ in Fig.1a should be accordingly modified.
The only other figure which is affected by such change is
Fig.3a. Its analysis now becomes more involved and we postpone it
(A. Brignole, L.E. Ib\'a\~nez, C. Mu\~noz and C. Scheich,
to appear).

\listrefs

{\centerline{\bf Figure Captions}}

Fig.1. Soft terms in units of the gravitino mass versus $\theta$.
a) C.Y. scenario; b) O-I scenario; c) O-II scenario.

Fig.2. Squark, slepton and gluino masses versus gluino mass.
a) C.Y. scenario; b) O-I scenario; c) O-II scenario.

Fig.3. Particle spectrum versus $\theta$ for a fixed gluino mass $M_g=500 GeV$.
a) C.Y. scenario, with $B=B_Z$; a') C.Y. scenario, with $B=B_{\mu}$;
b) O-I scenario; c) O-II scenario.

\end
\bye